\documentclass[twocolumn]{webofc}
\usepackage[varg]{txfonts}   % Web of Conferences font

\usepackage{subcaption}

\newcommand{\xm}{\relax\ifmmode X_{\mathrm{max}} \else
  $X_{\mathrm{max}}$\fi}
\newcommand{\mxm}{\relax\ifmmode \left<X_{\mathrm{max}}\right> \else
  $\left<X_{\mathrm{max}}\right>$\fi}
\newcommand{\sxm}{\relax\ifmmode \sigma(X_{\mathrm{max}}) \else
  $\sigma(X_{\mathrm{max}})$\fi}
\newcommand{\nm}{\relax\ifmmode N_{\mathrm{max}} \else
  $N_{\mathrm{max}}$\fi}

\begin{document}
\title{Measurements of UHECR Mass Composition by Telescope Array}
\author{\firstname{William}
  \lastname{Hanlon}\inst{1}\fnsep\thanks{\email{whanlon@cosmic.utah.edu}}
  for the Telescope Array Project}

\institute{University of Utah Dept. of Physics and Astronomy \& High
  Energy Astrophysics Institute, 201 James Fletcher Bldg., 115 S 1400
  E, Salt Lake City, UT 84112, USA}

\abstract{Telescope Array (TA) has recently published results of
  nearly nine years of \xm{} observations providing its highest
  statistics measurement of ultra high energy cosmic ray (UHECR) mass
  composition to date for energies exceeding $10^{18.2}$~eV. This
  analysis measured agreement of observed data with results expected
  for four different single elements. Instead of relying only on the
  first and second moments of \xm{} distributions, we employ a
  morphological test of agreement between data and Monte Carlo to
  allow for systematic uncertainties in data and in current UHECR
  hadronic models. Results of this latest analysis and implications of
  UHECR composition observed by TA are presented. TA can utilize
  different analysis methods to understand composition as both a
  crosscheck on results and as a tool to understand systematics
  affecting \xm{} measurements. The different analysis efforts
  underway at TA to understand composition are also discussed.}

\maketitle

\section{Introduction}\label{sec:intro}
Telescope Array (TA) is a large hybrid cosmic ray observatory located
in Millard County, Utah ($39.3^\circ$~N, $112.9^\circ$~W, 1400~m asl)
designed to observe ultra high energy cosmic rays with energies in
excess of $10^{18}$~eV. The addition of the TA Low Energy Extension
(TALE) has extended the minimum observable energy down to
$~10^{15.3}$~eV.

TA utilizes 507 plastic scintillation counters, also referred to as
surface detectors (SDs), placed over $~700$~km$^2$ and 36 fluorescence
detector (FD) telescopes to measure the energy, depth of air shower
maximum (\xm), and arrival direction of UHECRs. Three communications
towers are also deployed to allow wireless communications with the SDs
to facilitate readout of SD data for cosmic ray events and SD system
monitoring. Each SD consists of two layers of plastic scintillator,
each with area of 3~m$^2$ and 1.2~cm thick. Each layer has 104
wavelength shifting fiber optic cables embedded in them which are
optically coupled to a photomultiplier tube (PMT). Plastic
scintillators are sensitive to charged particles that arrive at ground
level as well as $\gamma$s, which is important in the case of UHECR
air showers since $\ge 90$~\% of the primary particle energy is stored
in the electromagnetic component of the shower. Each SD is equipped
with an electronics box, GPS antenna, solar panel, battery, and a
wireless local area network (WLAN) antenna. SDs are arranged in a
grid-like fashion with 1.2~km spacing between adjacent units. SDs
digitize signals from the PMTs via 12 bit flash analog-to-digital
(FADC) electronics with a 50~MHz sampling rate (20~ns time
resolution). A single waveform is 2.56~$\mu$s (128 FADC bins) long
and when event readout is ordered by one of the communication towers
up to 10 waveforms can be sent. Signals greater than 0.3 minimum
ionizing particles (MIP) generate a level-0 trigger, and signals
greater than 3.0 MIP generate a level-1 trigger. Communications towers
query all SDs each second for detection of level-1 triggers. When
three adjacent SDs record a level-1 trigger within 8~$\mu$s, a level-2
event trigger is generated. When a level-2 trigger is generated, the
communication towers collect all level-0 triggers recorded by the SDs
within $\pm 32 \mu$s of the event trigger time. The event data is
regularly transferred by wireless communications to a central data
facility located several kilometers away for offline analysis. Refer
to \cite{AbuZayyad:2012kk} for further details of TA's SD operations.

There are three FD stations at TA located on the periphery of the SD
array. Each station points towards the center of the SD array at a
central laser facility which is located 21~km away. The northern site,
Middle Drum (MD) FD station, employs 14 FD telescopes repurposed from
the HiRes experiment, each consisting of a 5.1~m$^2$ spherical mirror,
$16 \times 16$ PMT cluster, and electronics rack for event readout,
trigger logic, and communications with a central timing computer and
data acquisition (DAQ) unit to store data \cite{AbuZayyad:2000uu}. The
14 telescopes are arranged in a two ring configuration, with seven
telescopes viewing from $3^\circ$ to $17^\circ$ in elevation and seven
telescopes viewing from $17^\circ$ to $31^\circ$. Total azimuthal
coverage is 112$^\circ$. Each PMT observes $~1$~millisteradian solid
angle of the sky. Also located at the Middle Drum site are 10 FD
telescopes used for the TALE detector. At the southwest and southeast
corners of the SD array are the Long Ridge (LR) and Black Rock Mesa
(BR) FD stations. These stations are comprised of newly built FD
telescopes and electronics for the TA experiment. Each consists of 12
telescopes, utilizing 6.8~m$^2$ spherical mirrors, a $16 \times 16$
PMT cluster camera, and associated electronics and power racks. The
twelve mirrors at each of these stations are also arranged in a two
ring configuration, with six mirrors in ring 1 observing between
3$^\circ$ - 18$^\circ$ in elevation, and six mirrors in ring 2
observing between 18$^\circ$ - 33$^\circ$. The total azimuthal
coverage of each of these stations is 108$^\circ$
\cite{Tameda:2009zza,Tokuno:2012mi}. Middle Drum uses sample and hold
electronics readout, while Long Ridge and Black Rock Mesa use FADC
electronics providing 14 bit FADC samples at a 10~MHz rate. The
details of operation and triggering differ between Middle Drum and the
southern stations, but in general terms the telescopes look for
fluorescence light generated by a UHECR generated air shower. Time
coincidence and pattern recognition algorithms are employed by each
telescope's electronics to generate mirror triggers, which when
directed by a station-wide central computer, are collected to form
event triggers which may involve multiple mirrors. Event data is read
out from each mirror's electronics rack and stored for later offline
analysis. Because the PMTs are operated at very high bias, FDs only
operate on moonless nights under favorable weather conditions, thereby
limiting their duty cycle to $~10$\%. SDs, on the other hand, operate
continuously with 100\% duty cycle barring any maintenance issues.

Cosmic ray energy, arrival time, and direction can be measured
independently by either the SD array or FDs. FD reconstruction
requires measuring a shower-detector plane which is determined by
fitting the tube pointing directions and trigger times. This fit
provides the distance to the shower for any given pixel which observes
its crossing. An inverse Monte Carlo procedure is performed with the
fitted geometry to find the air shower profile that results in the
best agreement between simulated and observed tube signals. The energy
of these air showers can be very well measured because most of an air
shower's energy is transferred to an electromagnetic component
($e^\pm$, $\gamma$), the shower evolution is observed along many
degrees of the sky, and calorimetry of electromagnetic air showers is
well understood. The largest systematic uncertainty is the
atmosphere, which is considered to consist of molecular scattering and
aerosol scattering components. For the atmosphere to act as an ideal
calorimeter, the temperature and pressure profiles as a function of
height would be well known and would be free of aerosols. SD event
reconstruction proceeds by using the timing information of FADC
waveforms, the known geometry of each SD's placement, and assumptions
about the lateral evolution a UHECR induced air shower to determine
the spatial-temporal aspects of the primary particle. Monte Carlo
simulations are used to relate the observed signal and reconstructed
zenith angle to primary particle energy. This relationship is
dependent upon the hadronic model used. Finally an energy correction
is made based upon the observed relationship of $E_{\mathrm{FD}}$ and
$E_{\mathrm{SD}}$, which is the reconstructed energy of the same
events as observed by FD reconstruction and SD reconstruction
respectively.

To improve reconstructed parameters of the primary cosmic ray, hybrid
measurements can be performed, which combines the information of SD
and FD measurements for events that are simultaneously viewed by both
types of detectors. Stereo FD measurements which utilize event data
from two or more FD stations to improve UHECR measurement are also
possible. For purposes of performing a UHECR composition measurement,
which relates the primary particle energy and atmospheric depth of
shower maximum (\xm), these improvements to reconstruction are
desirable to make precision measurements, because the combined power
of multiple observations of a shower track greatly reduces the
uncertainty in \xm. Monocular FD measurements of \xm{} can typically
have resolution of $~70$~g/cm$^2$ or worse, while hybrid and stereo
measurements improve resolution to $~20$~g/cm$^2$ or better depending
on energy. Invoking the superposition principle, for a given energy
bin, \mxm{} of an ensemble of air showers generated by a single
chemical species with mass number $A$ is related to primary particle
mass as $\mxm \propto D \ln(E/A)$, where $E$ is the primary particle
energy and $D \equiv d\mxm/d\ln E$ is the elongation rate. For a fixed
energy, shower-to-shower fluctuations in \xm{} are large for a single
element. \sxm{} is roughly proportional to $\sigma_{p}/\sqrt{N}$,
where $\sigma_p$ is the standard deviation of the \xm{} distribution
of protons, and $N$ is the number of nucleons in the primary
particle. Effects such as multiplicity make this estimate more of a
lower bound. Mixtures of primary elements exhibit similar
relationships based upon $\left< \ln A \right>$. Utilizing these
relationships among the first and second moments of \xm{}
distributions, observed \xm{} can be used to measure UHECR
composition. The relationships between UHECR mass and \mxm{} and
\sxm{} can be summarized as \mxm{} and \sxm{} are larger for light
primaries.

The issue of model dependence must also be recognized as
a key ingredient in inferring composition from data when compared to
simulations. \xm{} and \sxm{} of hadronic showers are dependent upon
parameters such as multiplicity, inelasticity, and cross section, all
of which cannot be measured in the lab at the UHECR energy scale
($E_{\mathrm{lab}} \ge 10^{18}$~eV). Model uncertainties are a major
source of systematic uncertainty in measuring UHECR composition. See
\cite{Matthews:2005sd,Kampert:2012mx,Ulrich:2010rg} for detailed
expositions of theory of UHECR composition measured through
observations of air showers.

Telescope Array has performed four different measurements of composition:
two hybrid measurements, one stereo measurement, and one utilizing
only SDs. Section~\ref{sec:hybrid_fd} will discuss hybrid and FD
measurements, section~\ref{sec:sd} will discuss the SD measurement of
composition, and section~\ref{sec:summary} will summarize TA's
composition results.

\section{FD and Hybrid Composition Measurements}\label{sec:hybrid_fd}
\begin{figure*}
  \centering
  \begin{subfigure}{0.45\linewidth}
    \includegraphics[clip,width=\textwidth]{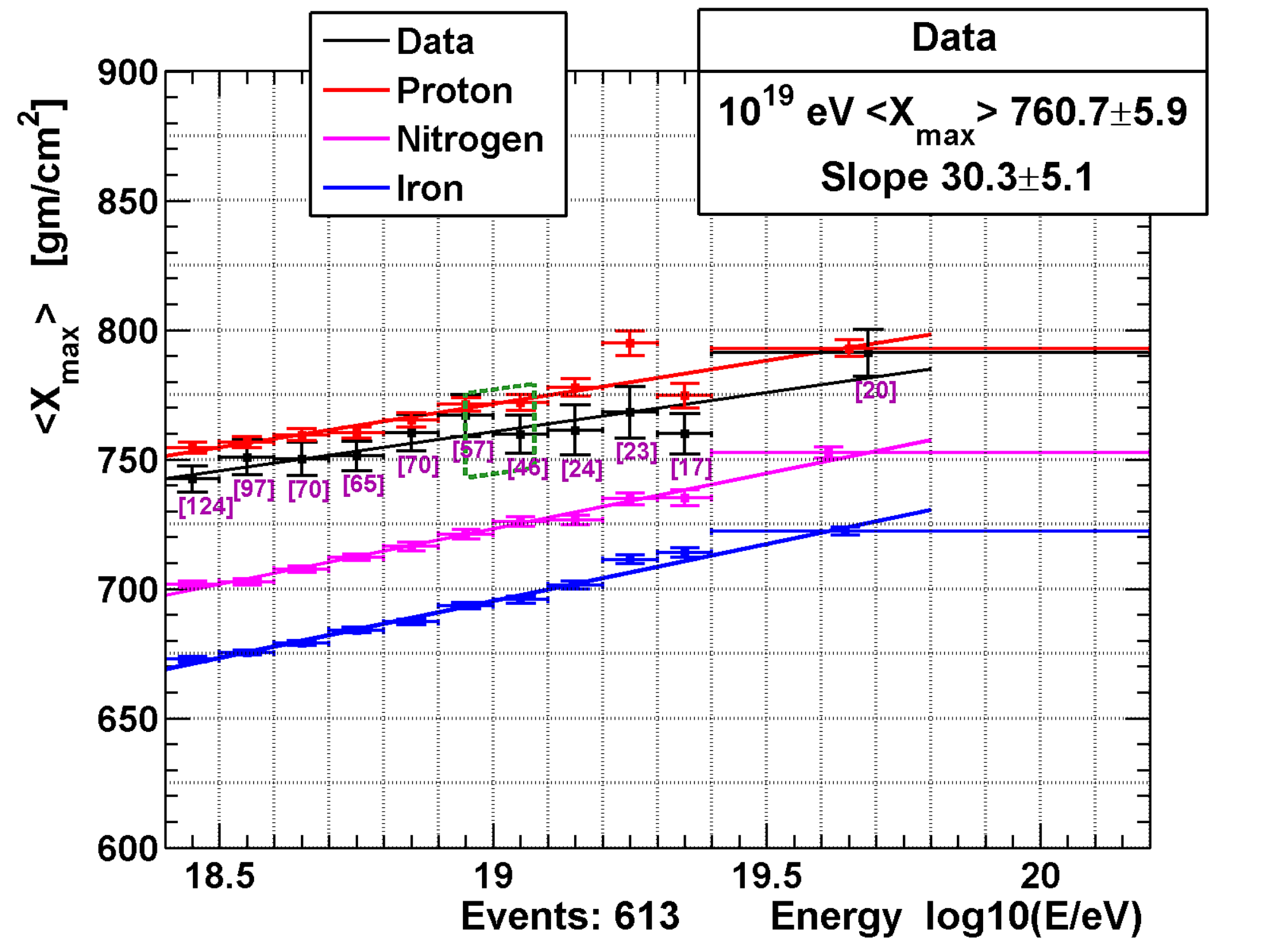}
    \caption{Observed \mxm{} of seven years of Middle Drum hybrid
    data.}
    \label{fig:md_mxm}
  \end{subfigure}%
  \qquad%
  \begin{subfigure}{0.45\linewidth}
    \includegraphics[clip,width=\textwidth]{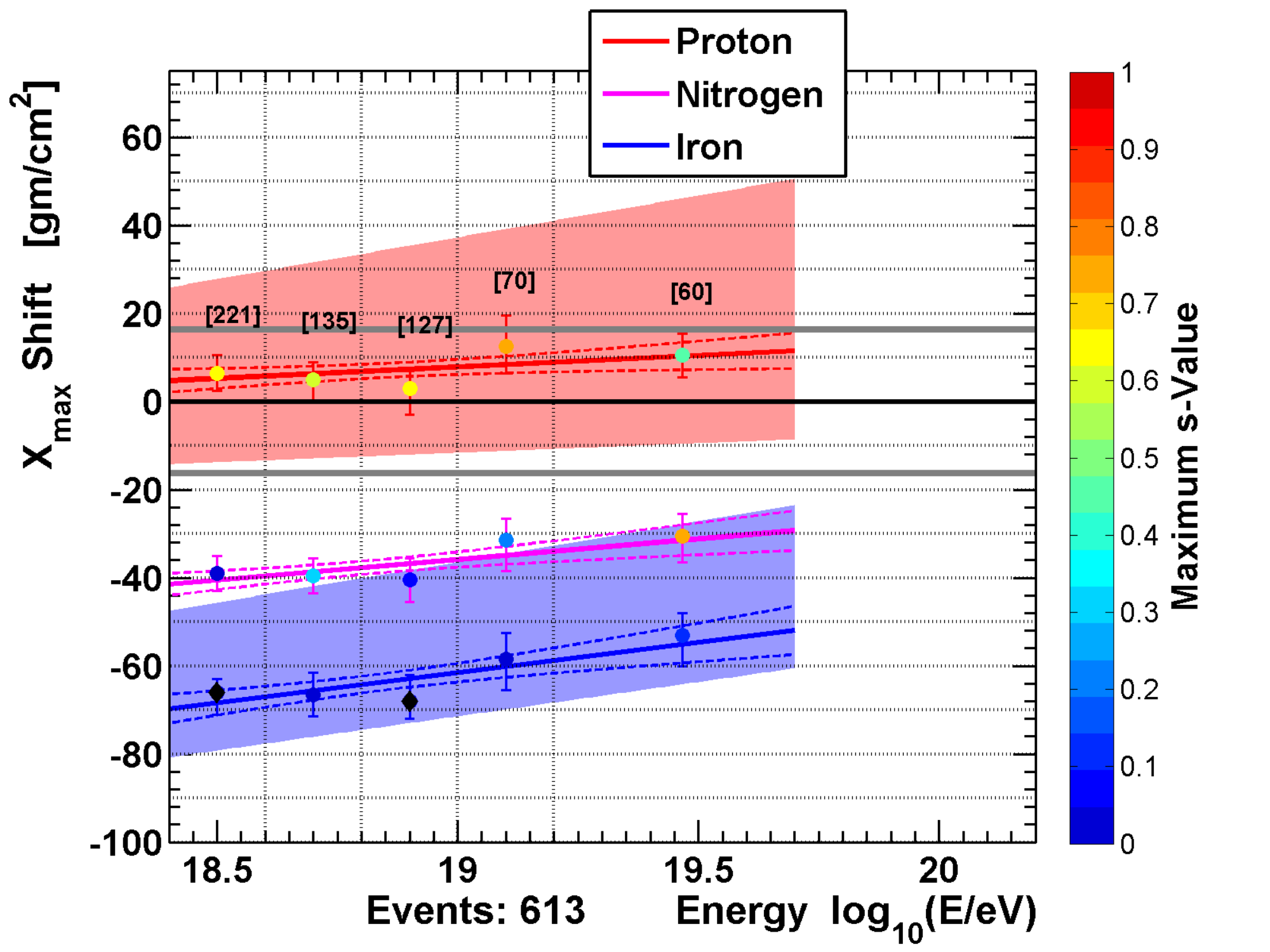}
    \caption{\mxm{} shift of Middle Drum data required to maximize
      Cram\'{e}r-von Mises test probability relative to predictions of
    proton, nitrogen, and iron.}
    \label{fig:md_cvm}
  \end{subfigure}
  \caption{Middle Drum \mxm{} and Cram\'{e}r-von Mises shift test. The
    data shows agreement with light composition.}
  \label{fig:md_xmax}
\end{figure*}

Two hybrid measurements of composition at TA have been done. One using
the Middle Drum FD station and a second using the Black Rock Mesa and
Long Ridge stations. There are sufficient differences in the design
and location of Middle Drum compared to Black Rock Mesa and Long Ridge
to necessitate analyzing their data separately. Middle Drum is placed
further from the SD array border than either Black Rock Mesa or Long
Ridge (8~km vs. 3~km and 4~km, respectively). Middle Drum also uses
smaller mirrors. These differences reduce the acceptance of Middle
Drum compared to Black Rock Mesa and Long Ridge, and the data analysis
for the two data sets are carried out independently.

Hybrid reconstruction is done by searching for coincident events in
the SD and FD data streams that occur within a small time window. The
timing and geometry of the event from the SD event data is used to
constrain the location of the shower core on the ground, which greatly
improves the determination of the shower track in the FD
shower-detector plane. Using the improved geometry fit, the light
profile of the shower is fit using the FD information, providing
accurate measurements of energy and \xm{} that are better than
monocular FD reconstruction alone. Uncertainties in angular
quantities important to reconstruction of the shower track improve to
less than a degree, and relative uncertainties in distances improve to
less than 1\% when performing hybrid reconstruction.

Results of five years of Middle Drum hybrid data have been published
in 2015~\cite{Abbasi:2014sfa} and extended to seven years of analysis
in 2016~\cite{Lundquist:2015cgq}. Results and plots presented here are
from the seven year analysis covering the period from May 31, 2008 to
April 24, 2015, which resulted in 613 events collected. Using the
hybrid reconstructed events observed by Middle Drum a pattern
recognition algorithm is applied to ensure the shower profile is well
behaved and the rise and fall of the shower is in the FD field of
view. Events that pass the pattern recognition step have further
quality cuts applied: good weather cuts to remove clouds in FD field
of view, $E > 10^{18.4}$~eV, zenith angle $< 58^\circ$, shower core
not further than 500~m outside the SD array boundary, SD/FD core
difference $< 1600$~m, geometry $\chi^2$/DOF < 5.

Figure~\ref{fig:md_mxm} shows observed \mxm{} of the seven year Middle
Data analysis. Red, magenta, and blue points show the predictions of
QGSJET~II-03 proton, nitrogen, and iron. The green box indicates the
systematic uncertainty on \mxm{} for this analysis, calculated to be
16~g/cm$^2$. The reconstructed \xm{} bias for $E > 10^{18.4}$~eV is $<
2$~g/cm$^{2}$, and resolution is 22~g/cm$^2$ for this analysis. TA
Middle Drum \mxm{} appears to be consistent with a predominantly light
composition over the entire energy range shown in the figure. Current
generation UHECR observatories have sufficient exposure to collect
large event samples. Because \xm{} distributions are naturally skewed,
especially in the case of light primaries, the first and second
moments of these distributions may obscure some valuable information
related to composition, namely if a deep \xm{} tail is present. If
there is a significant contribution of light primaries their deeply
penetrating nature will be revealed by the presence of a tail in the
\xm{} distribution. A better way to test agreement between data and
models is to use the full \xm{} distributions instead of simply \mxm{}
and \sxm{}. Some statistical tests which may be useful are
distribution-free, two sample tests such Anderson-Darling,
Kolmogorov-Smirnov, and Cram\'{e}r-von Mises. Traditional $\chi^2$ and
maximum likelihood hypothesis testing can also be used to calculate
$p$-values to test the assumption that observed data is compatible
with models. Because of the skewed nature of \xm{} distributions, the
Anderson-Darling or Cram\'{e}r-von Mises test, which are quadratic
empirical distribution function (EDF) tests that measure the
integrated squared difference between two EDFs, are better suited for
tests of compatibility between observed and simulated \xm{}
distributions. The Kolmogorov-Smirnov test calculates the supremum of
the EDFs under comparison and is less sensitive to differences in the
tails of the distributions~\cite{Porter:2008mc}.

To test the compatibility of observed Middle Drum \xm{} with single
species simulated using the QGSJET~II-03 hadronic model, the
Cram\'{e}r-von Mises test was used. The data distributions were
uniformly shifted to find the $\Delta \xm$ shift which provided the
best test statistic, and the $p$-value was calculated for the
statistic. Because one of the distributions was allowed to shift when
calculating the test statistic, it is referred to as a shape value, or
$s$-value. The interpretation of the $s$-value will be the same
though, if the $s$-value is large, then the probability that data and
the simulated \xm{} distributions are drawn from the same parent
distribution is more likely. Figure~\ref{fig:md_cvm} shows the \mxm{}
shift of data required to maximize agreement of the Cram\'{e}r-von
Mises test statistic with Monte Carlo predictions for QGSJET~II-03
proton, nitrogen, and iron. Color of the data points indicate the
$s$-value of the test. Colored bands show the range of shifts obtained
for the same test for several different models. Large $s$-values with
relatively small shifts are observed for QGSJET~II-03 protons
indicating Middle Drum \xm{} data is more likely compatible with
protons than either nitrogen or iron.

Stereo FD reconstruction is similar to hybrid except SD information is
not used. Instead, events that are observed simultaneously by multiple
FD stations are utilized. By using the intersecting shower-detector
planes to constrain the parameters of the shower track observed by two
or possibly three FD stations, the shower geometry can be very well
measured. Because of the relatively large separation between FD
stations at TA, $~30 - 40$~km, the stereo aperture is smaller than
hybrid at low energies. FD reconstruction allows for larger zenith
angle acceptance though, because SD reconstruction is limited to
$\theta < 60^\circ$ due to large uncertainties in the shower footprint
at such large zenith angles. TA's FD analysis can reconstruct shows
with zenith angles up to $80^\circ$. Therefore at high energies the
stereo aperture continues to grow because higher energy showers must
be inclined sufficiently to ensure \xm{} occurs in the air and not in
the ground. These steep zenith angle events are not reconstructed in
hybrid analyses. For the stereo analysis results presented here, the
minimum accepted energy is $10^{18.4}$~eV.

\begin{figure}
  \centering
  \includegraphics[clip,width=0.98\linewidth]{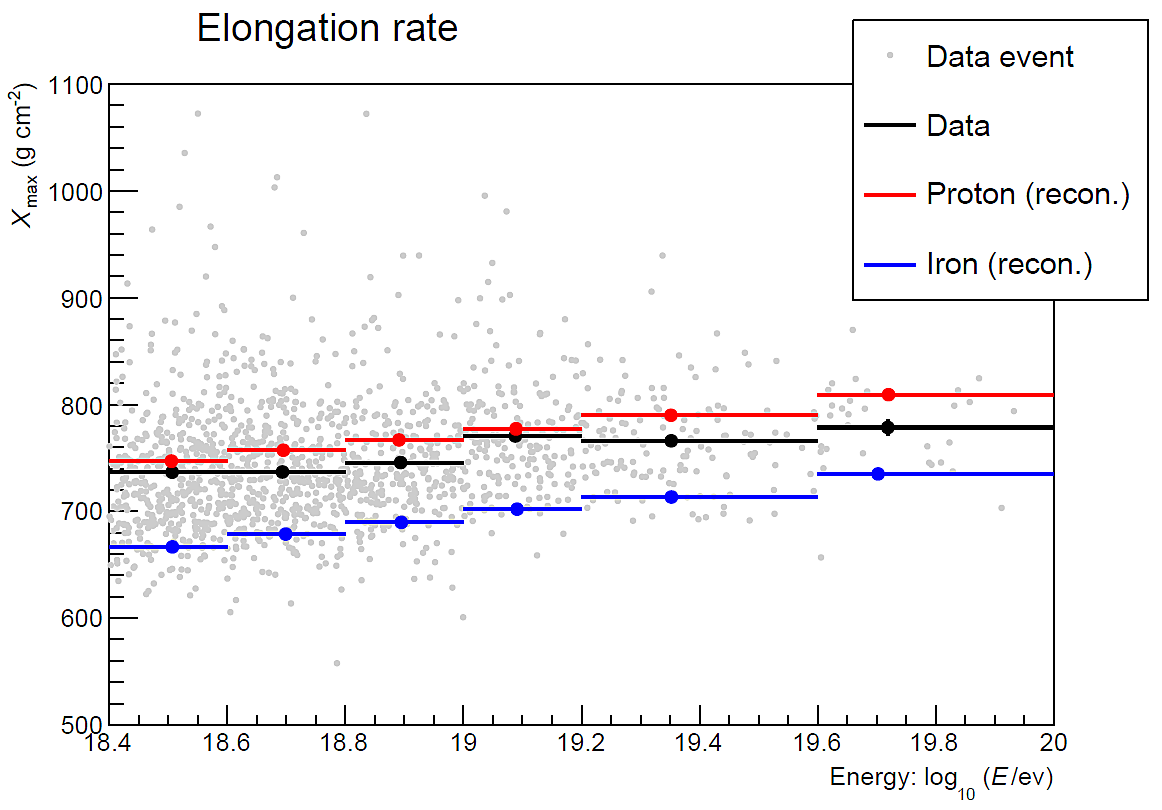}
  \caption{Nine year TA stereo \mxm{} compared to QGSJET~II-03
    expectation. The observed data is consistent with light composition.}
  \label{fig:stereo_mxm}
\end{figure}

Figure~\ref{fig:stereo_mxm} shows the observed stereo \mxm{} collected
using nine years of data resulting in 1458 events collected, as well
as the predictions for QGSJET~II-03 protons and iron. The systematic
uncertainty of the data is 15~g/cm$^2$. \xm{} bias and resolution is
0.9 and 19.2~g/cm$^2$ respectively for QGSJET~II-03 protons, and
energy resolution is 5.1\%~\cite{Bergman:2017ikd}. The observed data
is consistent with a predominantly light composition.

\begin{figure}
  \centering
  \includegraphics[clip,width=0.94\linewidth]{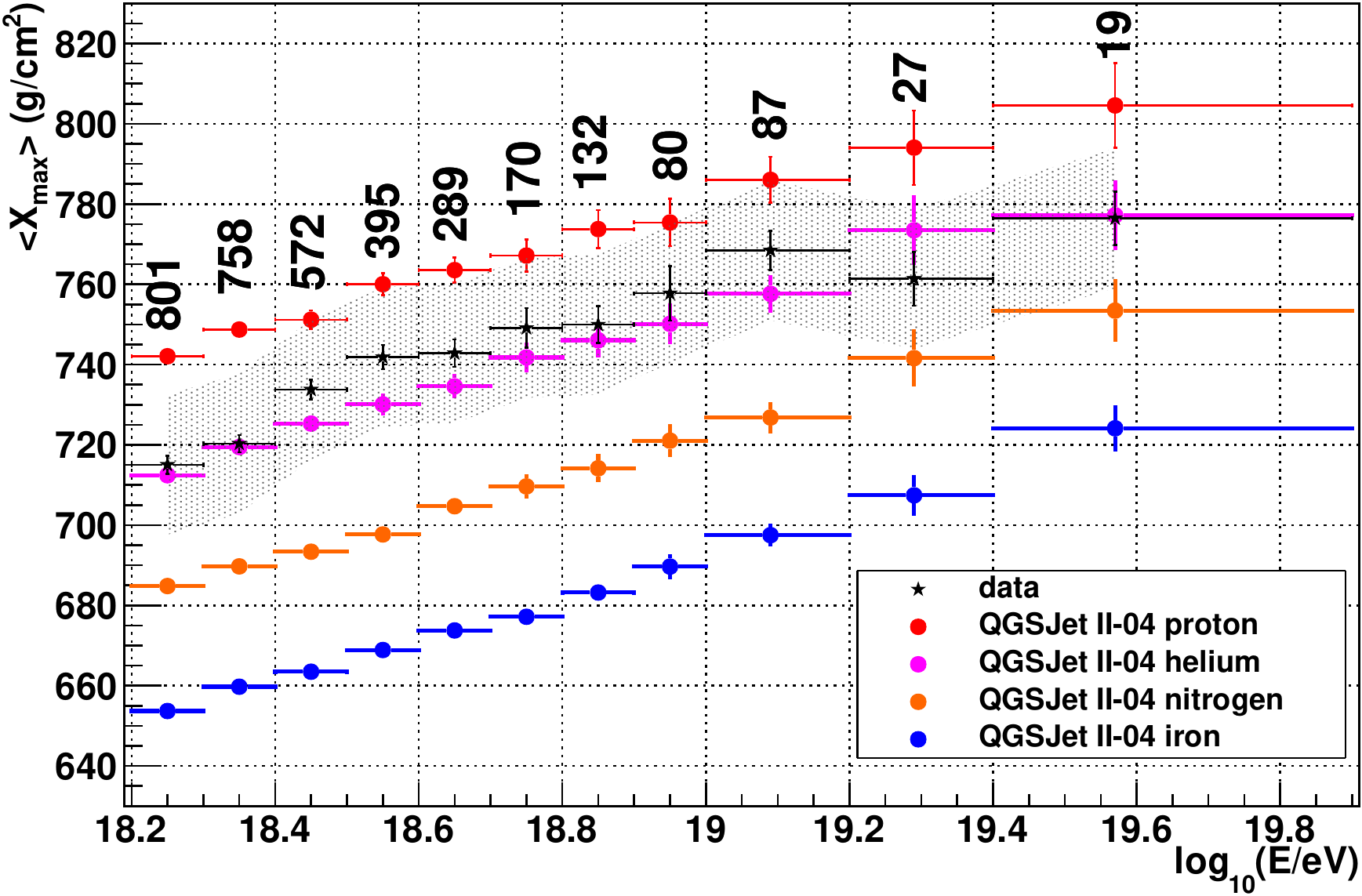}
  \caption{BR/LR hybrid observed and predicted \mxm{} for four
    different chemical elements. The number of data events for energy
    bin are also shown.}
  \label{fig:brlr_mxm}
\end{figure}

\begin{figure*}
  \begin{subfigure}{0.46\linewidth}
    \includegraphics[clip,width=\linewidth]{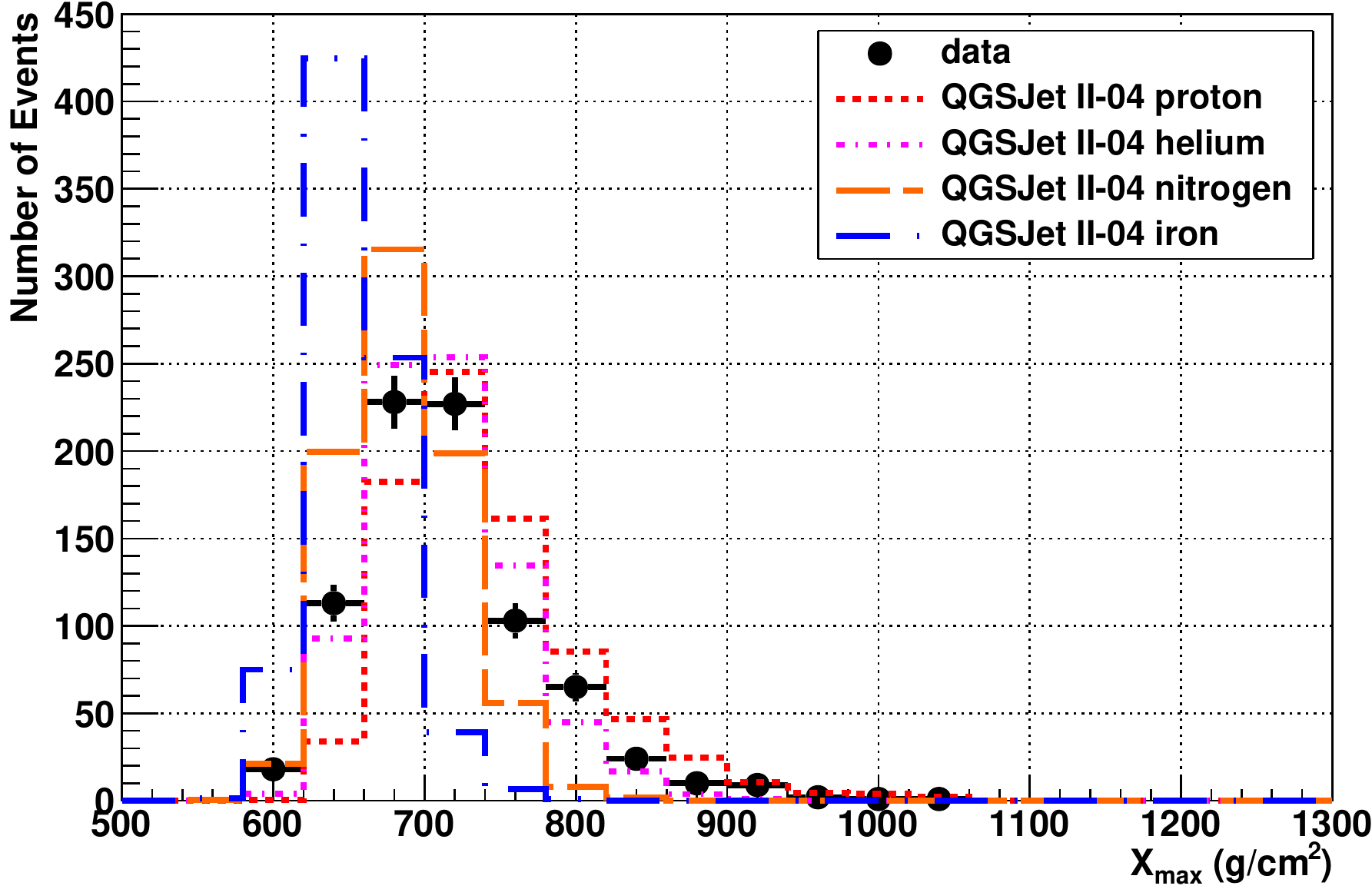}
    \caption{$18.2 \leq \log_{10}(E/\mathrm{eV}) < 18.3$}
    \label{fig:brlr_data_mc_comp_00}
  \end{subfigure}%
  \qquad%
  \begin{subfigure}{0.46\linewidth}
    \includegraphics[clip,width=\linewidth]{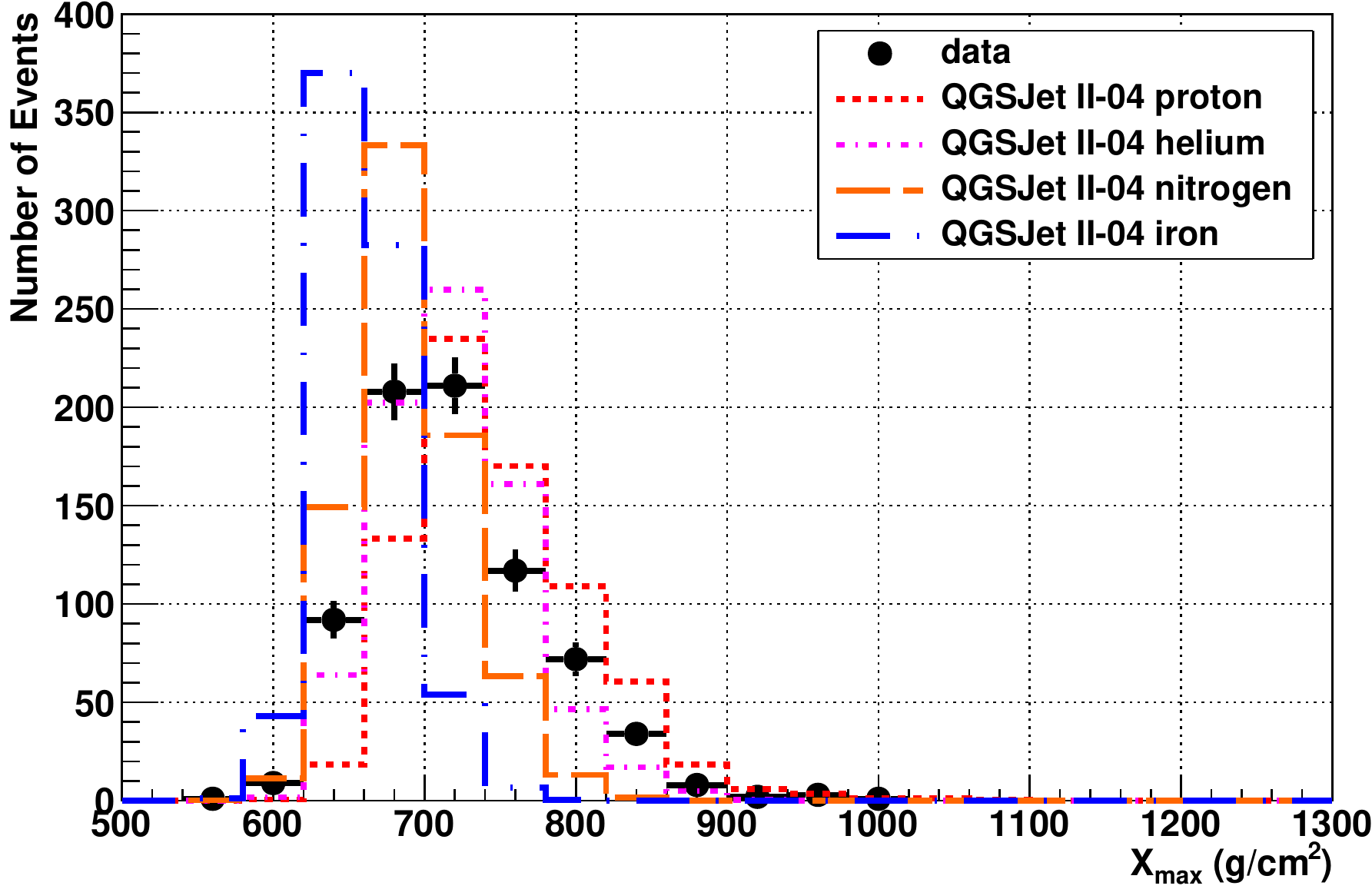}
    \caption{$18.3 \leq \log_{10}(E/\mathrm{eV}) < 18.4$}
    \label{fig:brlr_data_mc_comp_01}
  \end{subfigure}

  \begin{subfigure}{0.46\linewidth}
    \includegraphics[clip,width=\linewidth]{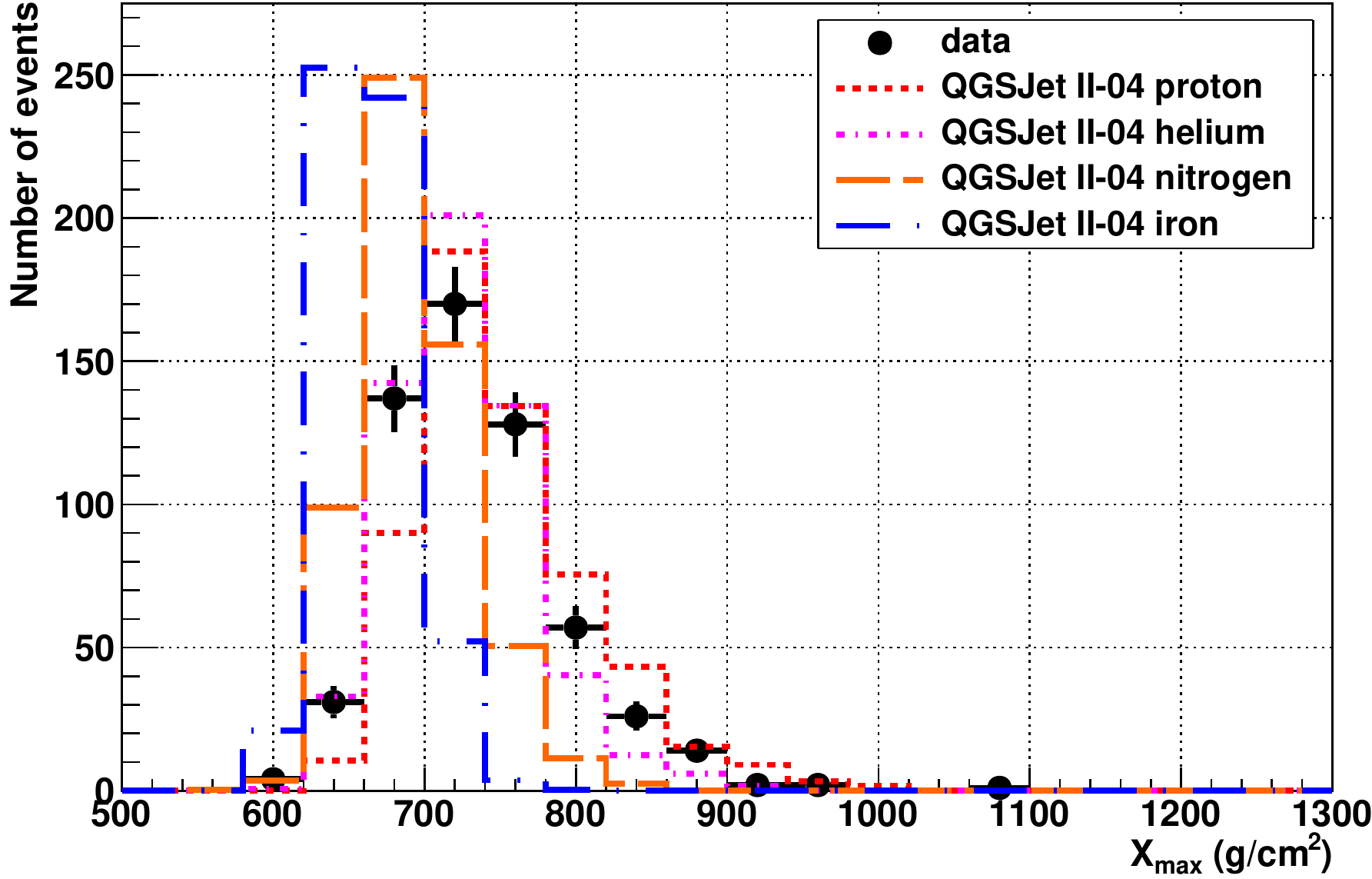}
    \caption{$18.4 \leq \log_{10}(E/\mathrm{eV}) < 18.5$}
    \label{fig:brlr_data_mc_comp_02}
  \end{subfigure}%
  \qquad%
  \begin{subfigure}{0.46\linewidth}
    \includegraphics[clip,width=\linewidth]{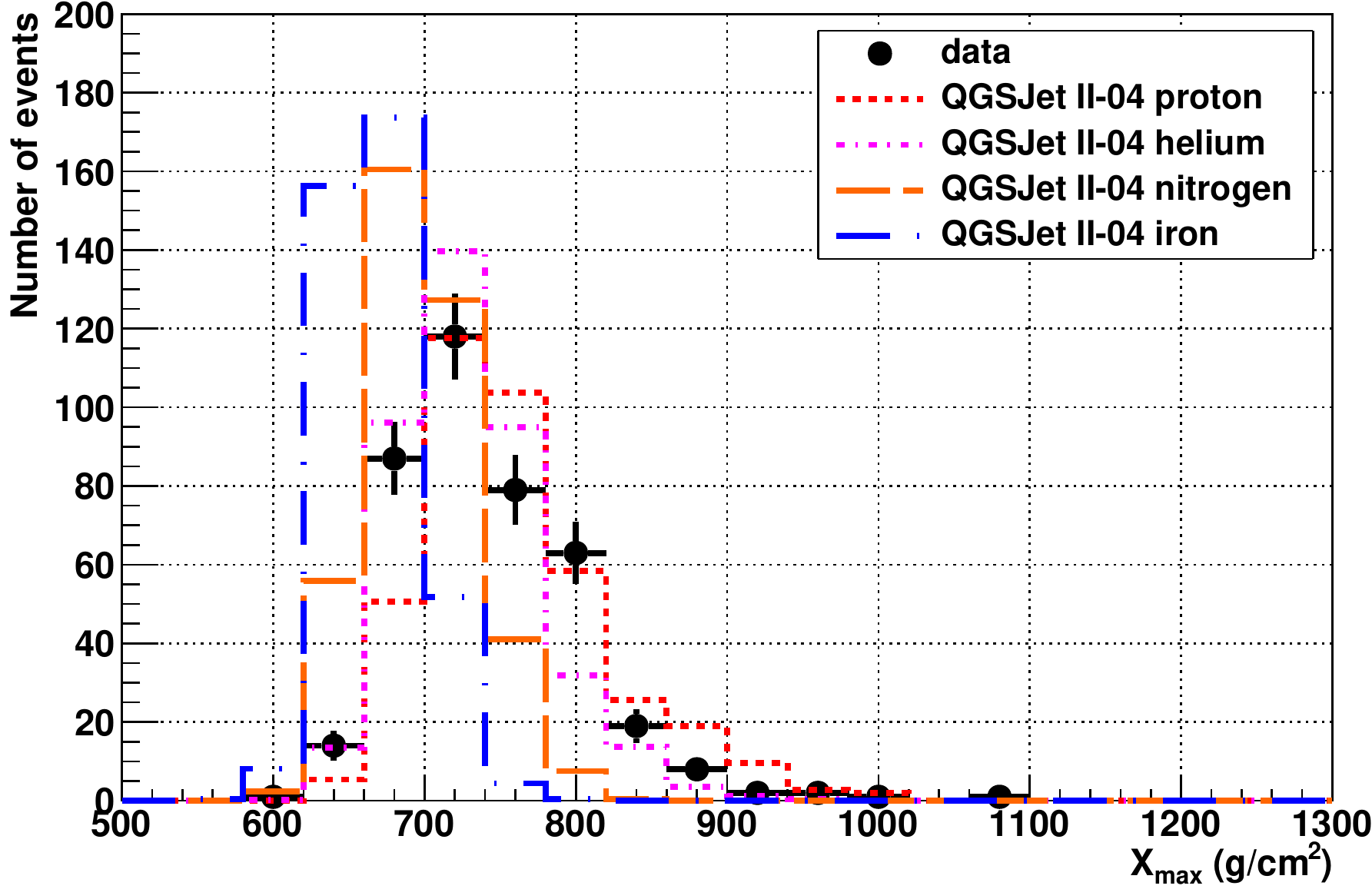}
    \caption{$18.5 \leq \log_{10}(E/\mathrm{eV}) < 18.6$}
    \label{fig:brlr_data_mc_comp_03}
  \end{subfigure}

  \begin{subfigure}{0.46\linewidth}
    \includegraphics[clip,width=\linewidth]{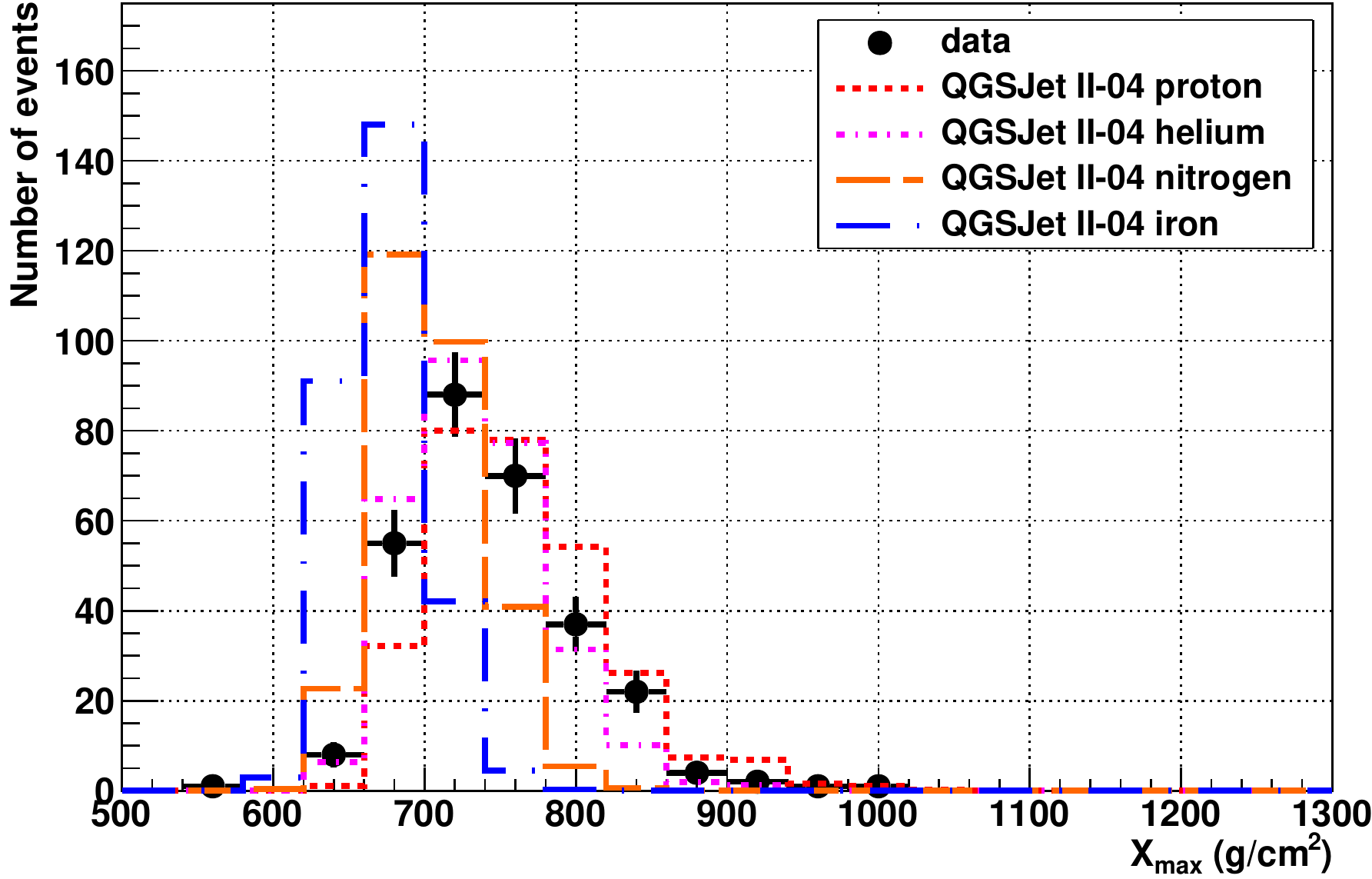}
    \caption{$18.6 \leq \log_{10}(E/\mathrm{eV}) < 18.7$}
    \label{fig:brlr_data_mc_comp_04}
  \end{subfigure}%
  \qquad%
  \begin{subfigure}{0.46\linewidth}
    \includegraphics[clip,width=\linewidth]{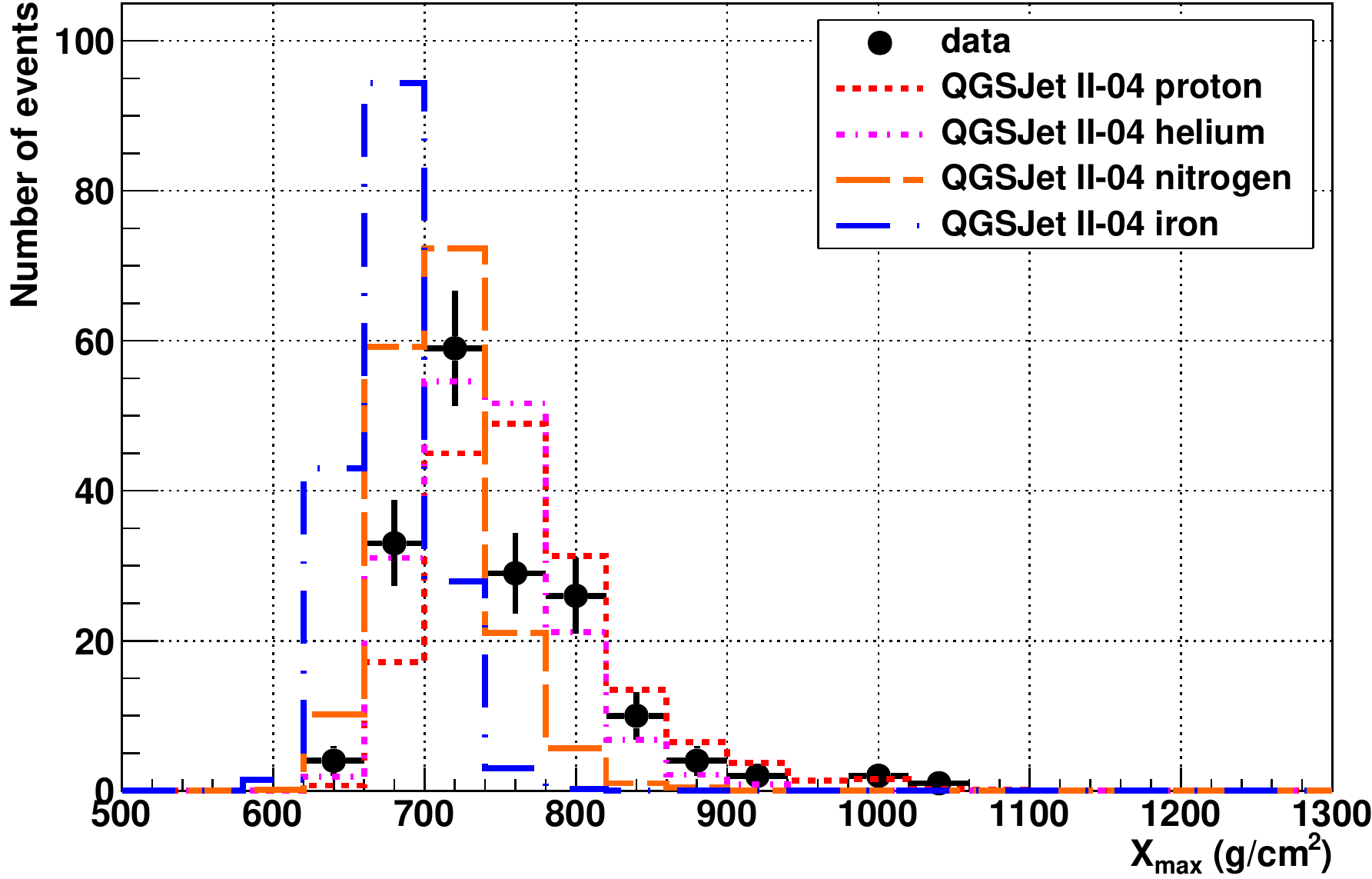}
    \caption{$18.7 \leq \log_{10}(E/\mathrm{eV}) < 18.8$}
    \label{fig:brlr_data_mc_comp_05}
  \end{subfigure}

  \begin{subfigure}{0.46\linewidth}
    \includegraphics[clip,width=\linewidth]{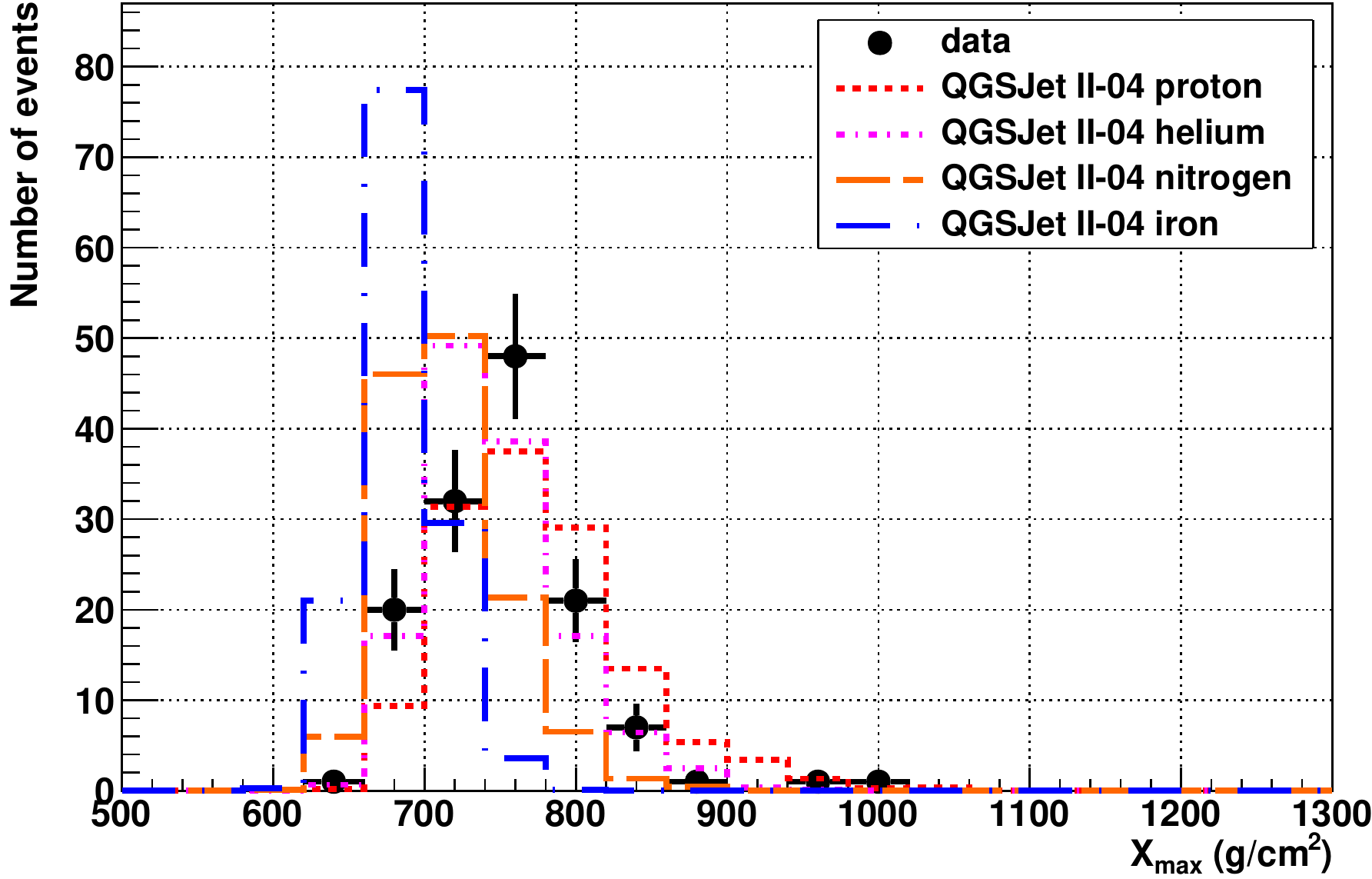}
    \caption{$18.8 \leq \log_{10}(E/\mathrm{eV}) < 18.9$}
    \label{fig:brlr_data_mc_comp_06}
  \end{subfigure}%
  \qquad%
  \begin{subfigure}{0.46\linewidth}
    \includegraphics[clip,width=\linewidth]{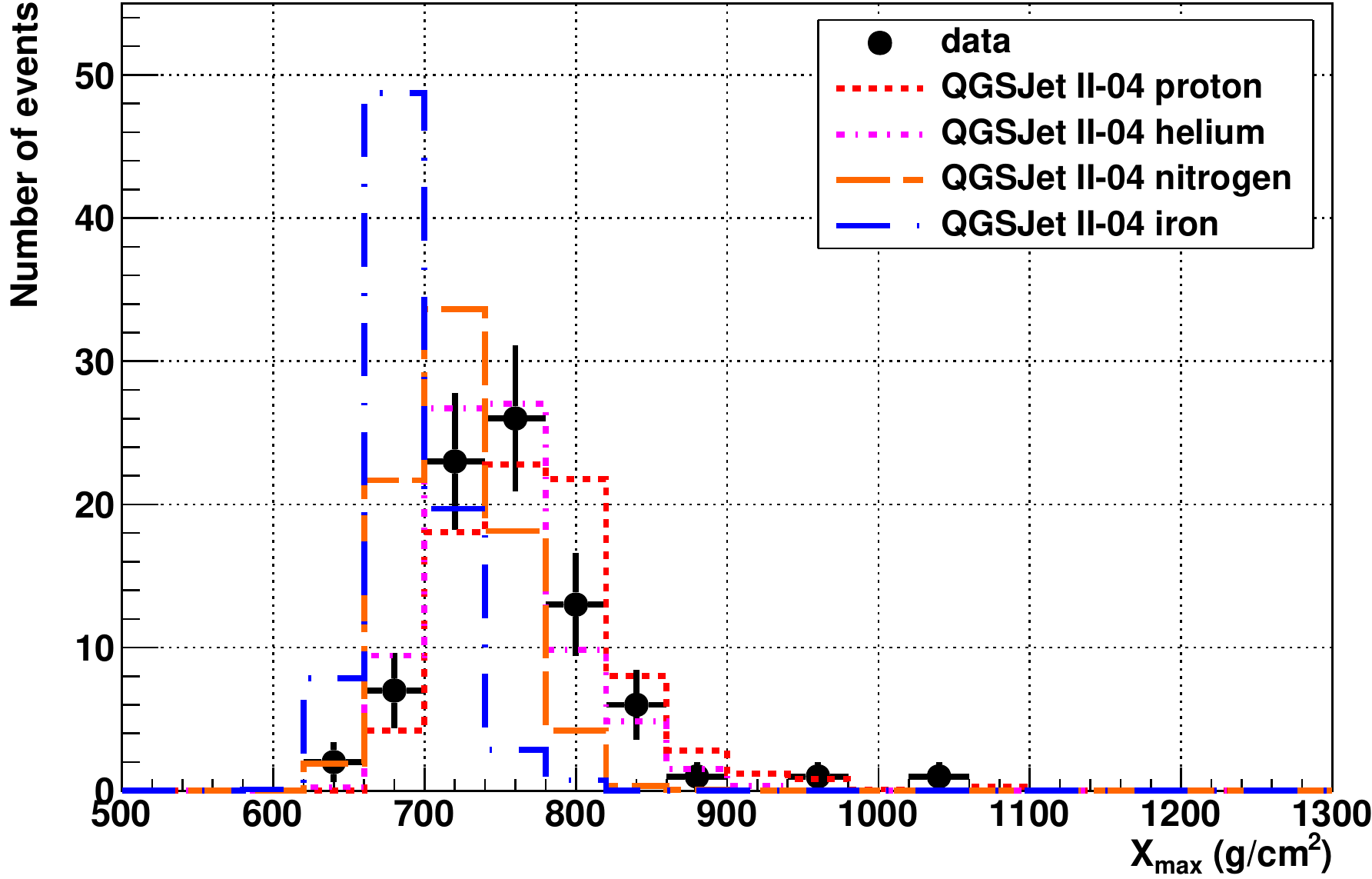}
    \caption{$18.9 \leq \log_{10}(E/\mathrm{eV}) < 19.0$}
    \label{fig:brlr_data_mc_comp_07}
  \end{subfigure}
  \caption{BR/LR hybrid \xm{} distributions I.}
  \label{fig:brlr_data_mc_comp_i}
\end{figure*}

\begin{figure*}
  \begin{subfigure}{0.46\linewidth}
    \includegraphics[clip,width=\linewidth]{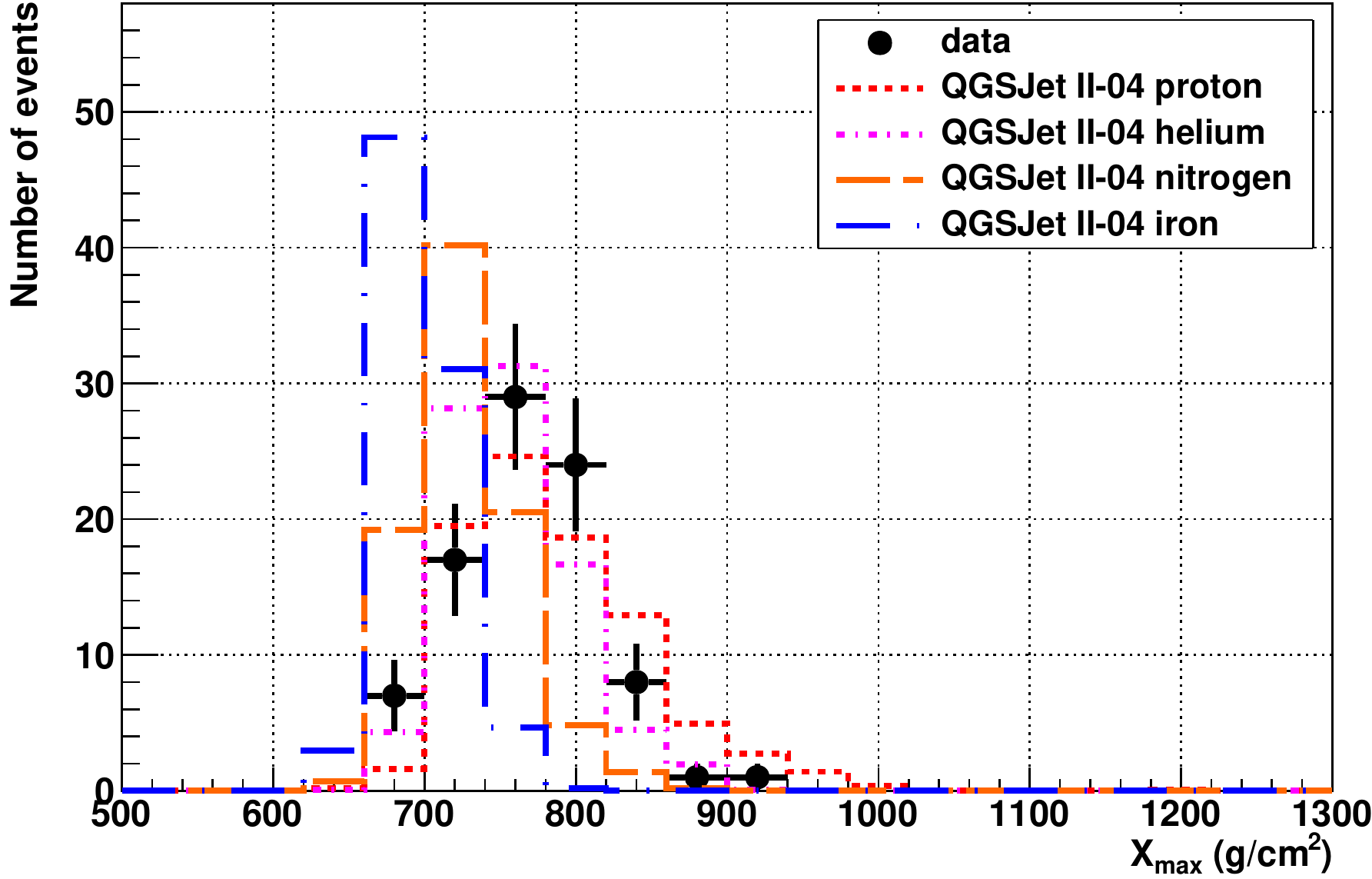}
    \caption{$19.0 \leq \log_{10}(E/\mathrm{eV}) < 19.2$}
    \label{fig:brlr_data_mc_comp_08}
  \end{subfigure}%
  \qquad%
  \begin{subfigure}{0.46\linewidth}
    \includegraphics[clip,width=\linewidth]{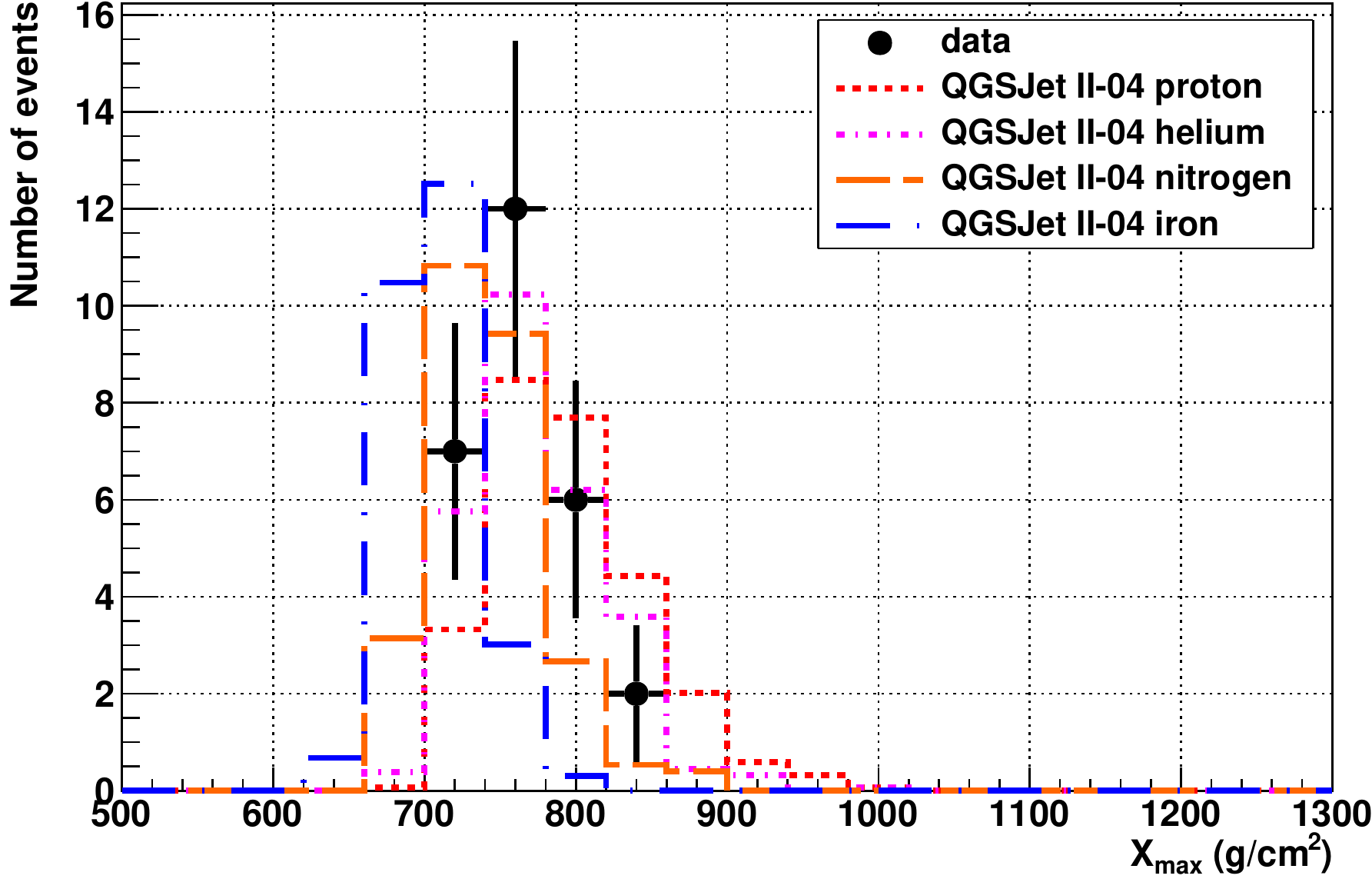}
    \caption{$19.2 \leq \log_{10}(E/\mathrm{eV}) < 19.4$}
    \label{fig:brlr_data_mc_comp_09}
  \end{subfigure}

  \begin{subfigure}{0.46\linewidth}
    \includegraphics[clip,width=\linewidth]{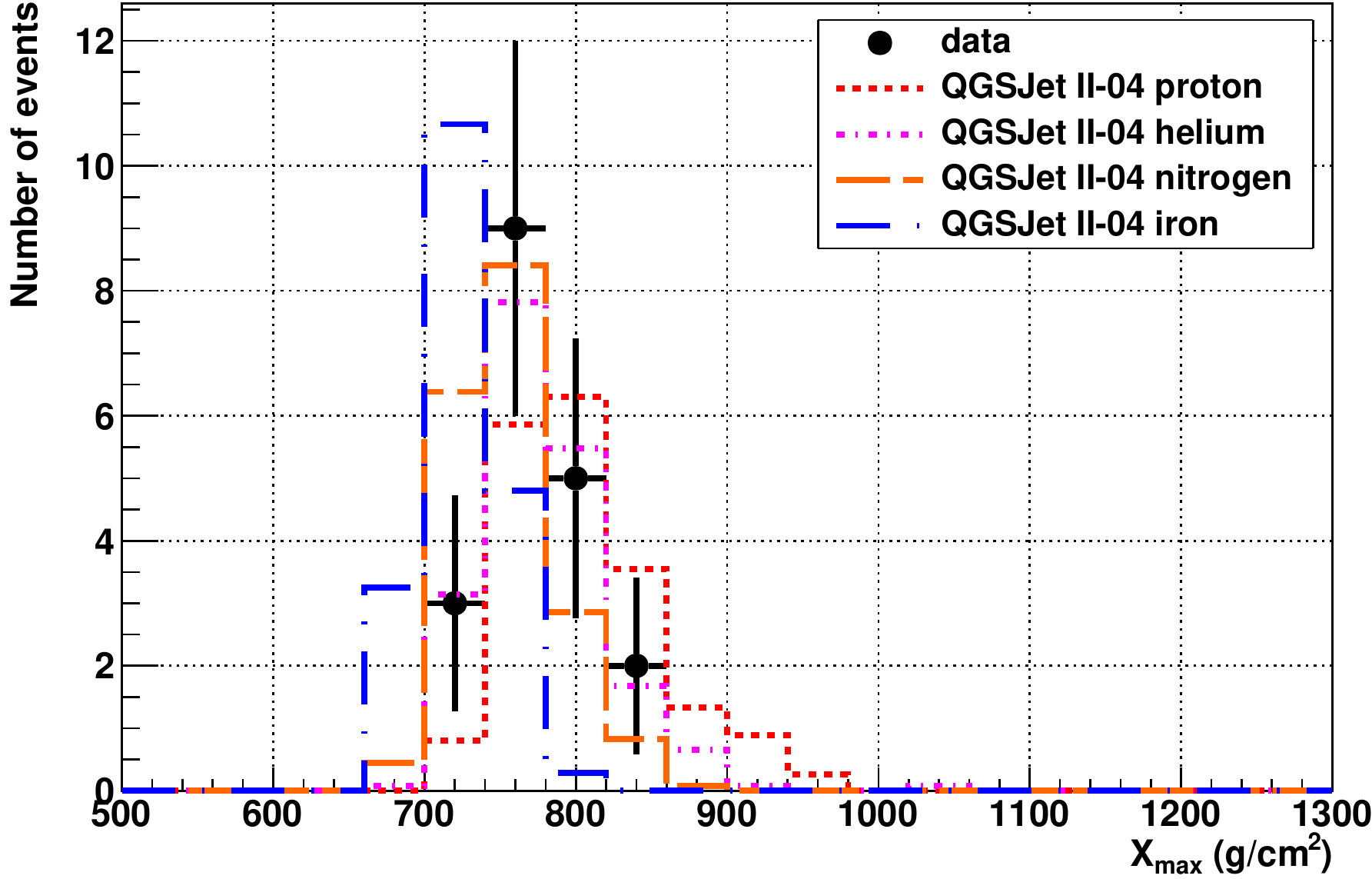}
    \caption{$19.4 \leq \log_{10}(E/\mathrm{eV}) < 19.9$}
    \label{fig:brlr_data_mc_comp_10}
  \end{subfigure}
  
  \caption{BR/LR hybrid \xm{} distributions II.}
  \label{fig:brlr_data_mc_comp_ii}
\end{figure*}

TA's highest statistics measure of composition is done using Black
Rock Mesa and Long Ridge (BR/LR) hybrid. Events that trigger the BR or
LR FD stations are time matched to events that also trigger the SD
array. If an event is observed by both FD stations, the shower
parameters from the site with the better hybrid shower profile are
chosen. Figure~\ref{fig:brlr_mxm} shows the observed \mxm{} along with
predictions of QGSJET~II-04 proton, helium, nitrogen, and iron for
nearly 9 years of data. Data and Monte Carlo are processed via the
same analysis software and the same quality cuts are applied: the
event core must be greater than 100~m from the SD array boundary, FD track
length $10^\circ$ or greater, 11 or more good tubes recorded by FDs,
shower-detector plane angle less than $130^\circ$, time extent of the
FD track greater than 7~$\mu$s, zenith angle less than $55^\circ$,
\xm{} must be observed, and weather cuts to ensure atmospheric quality
is good. 3330 data events were reconstructed after application of
these cuts. The systematic uncertainty on the data is 17~g/cm$^{2}$
(black band in the figure). \xm{} bias and resolution are -1.1 and
17.2~g/cm$^2$ respectively, and energy resolution is
5.7\%~\cite{Abbasi:2018nun}. Figures~\ref{fig:brlr_data_mc_comp_i} and
\ref{fig:brlr_data_mc_comp_ii} show the observed and simulated \xm{}
distributions for each energy bin.

\begin{figure}
  \centering
  \includegraphics[clip,width=0.98\linewidth]{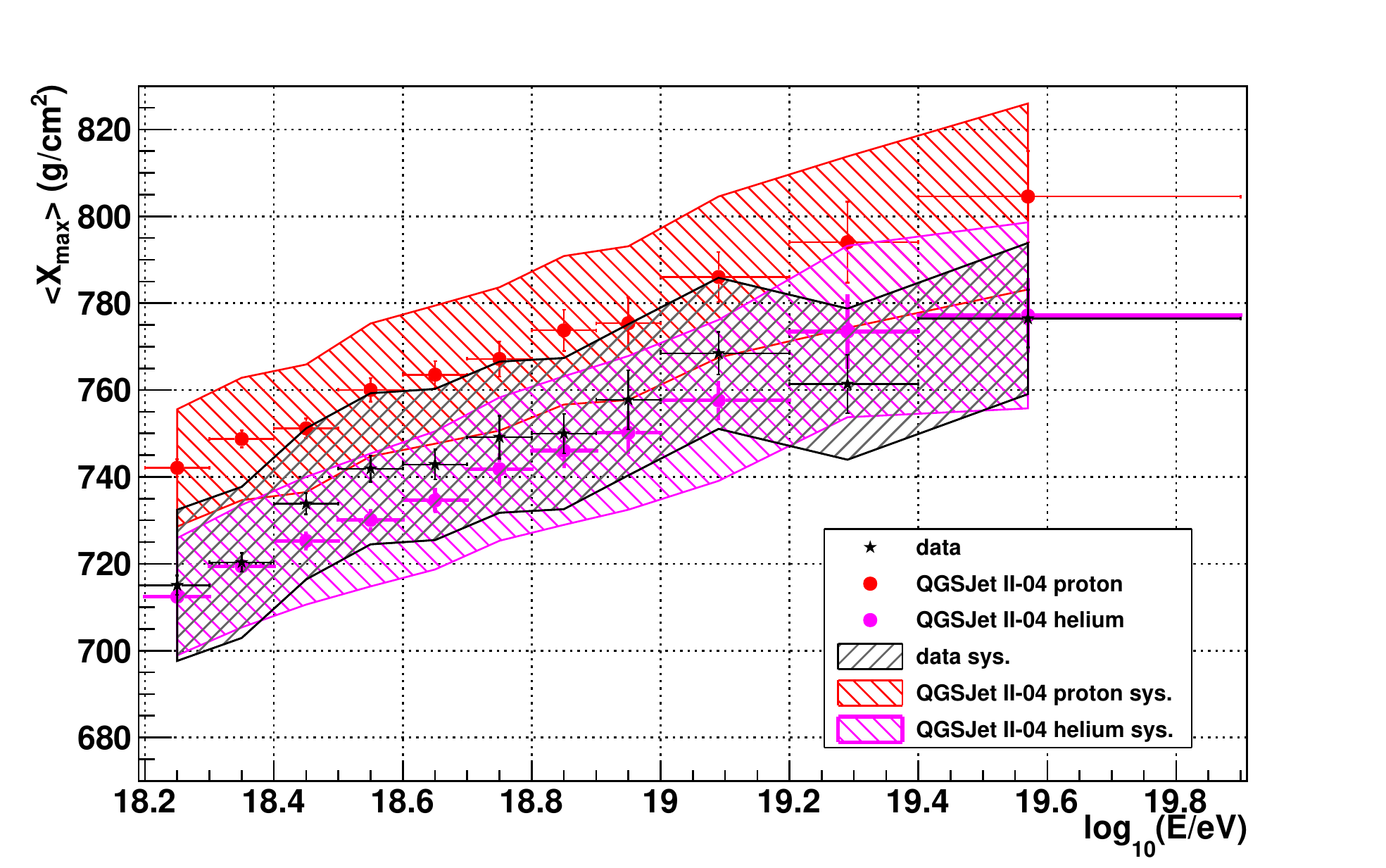}
  \caption{TA BR/LR hybrid \mxm{} compared to systematic uncertainty
    of the QGSJET~II-04 model predictions for proton and helium
    primary sources. Model systematics are estimated from
    \cite{Abbasi:2016sfu}.}
  \label{fig:model_sys}
\end{figure}

In addition to systematic uncertainty in event reconstruction and
detector acceptance, systematic uncertainty in hadronic models is a
major source of uncertainty when trying to answer the question of how
well data compares to them. This is because hadronic models require
knowledge of energy dependent quantities such as multiplicity,
inelasticity, and cross section to properly describe air shower
evolution. These parameters must be extrapolated from values which are
measured at relatively low energy and small pseudorapidity in collider
experiments. The Large Hadron Collider (LHC) currently reaches
$\sqrt{s} = 13$~TeV which is equivalent to $~10^{17}$~eV in the lab
frame, an one to three orders of magnitude below the typical event
energy analyzed in the results presented here. Additionally, collider
experiments typically observe processes with high transverse momentum
and small pseudorapidity, whereas air shower development is driven by
very high energy forward scattering processes. Current experiments at
LHC such as LHCf~\cite{Adriani:2008zz} and TOTEM~\cite{Anelli:2008zza}
are designed to probe physics in the high pseudorapidity
regime. Ulrich, Engel, and Unger examined the effect on various air
shower observables such as \nm{}, RMS(\xm), and EM fraction by varying
multiplicity, inelasticity, and cross
section~\cite{Ulrich:2010rg}. They found that RMS(\xm) is mainly
dependent on tuning of the cross section parameter and weakly
dependent on elasticity. Abbasi and Thomson extended this work to
examine the systematic uncertainty of \mxm{} for several different
hadronic models over a wide range of energies and estimated the
uncertainty in \mxm{} of QGSJET~II-04 to range from $\sigma(\mxm) =
\pm 3$~g/cm$^2$ at $10^{17}$~eV to $\pm 18$~g/cm$^2$ at
$10^{19.5}$~eV~\cite{Abbasi:2016sfu}). Figure~\ref{fig:model_sys} shows
the expected band of systematic uncertainty of TA BR/LR hybrid data,
as well as QGSJET~II-04 proton and helium predictions. The wide range
of systematic uncertainty in the models provides justification for not
merely relying upon \mxm{} and \sxm{} to interpret UHECR data, but
also testing the entire \xm{} distributions with \xm{} shifting.

\begin{figure}
  \centering
  \includegraphics[clip,width=0.98\linewidth]{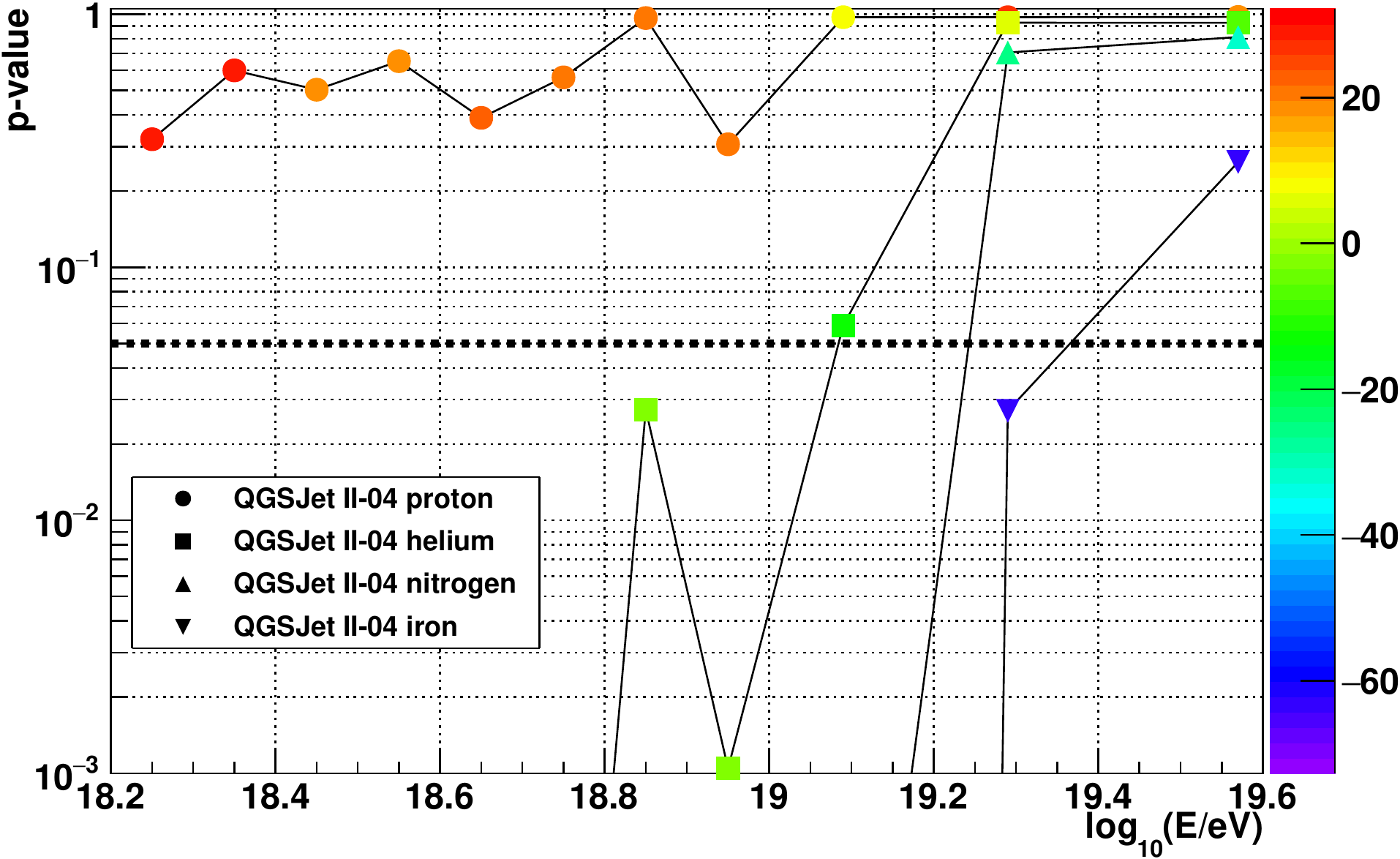}
  \caption{BR/LR hybrid \xm{} compatibility with different simulated
    elements}
  \label{fig:brlr_hypothesis_test}
\end{figure}

The BR/LR hybrid analysis also implemented a test which compares the
entire \xm{} distribution shapes, as was done for the Middle Drum
hybrid analysis. Here the maximum likelihood of the data \xm{}
distribution given the expected \xm{} distribution of proton, helium,
nitrogen, or iron in each energy bin was computed while allowing the
data distribution to shift systematically. The shift which provided
the best likelihood value was recorded as well as the $p$-value of
observing a likelihood at least as
extreme. Figure~\ref{fig:brlr_hypothesis_test} shows the results of
these tests. The ordinate of the plot indicates the probability after
systematic shifting of observing a likelihood at least as large as
that observed in the data given that the distribution was generated
from one of the pure chemical elements under test here. Small
$p$-values indicate disagreement with data and the model and is deemed
incompatible with the data. Large $p$-values indicate that the model
cannot be rejected as being compatible with the data. The color of
each point indicates the amount of systematic shifting that was
required to measure the observed $p$-value. In the last energy bin for
example, iron is shown to be compatible with data because of the large
$p$-value obtained, but a $~60$~g/cm$^2$ shift is required. This shift
is much larger than the systematic uncertainty of this analysis, and
therefore is should be considered skeptically. As
figures~\ref{fig:brlr_data_mc_comp_i} and
\ref{fig:brlr_data_mc_comp_ii} show the \xm{} distributions obtained
in the last few high energy bins show a marked suppression in the deep
\xm{} tail. This lack of the tail feature, which can act as a
discriminator of light elements in the \xm{} distribution, allows for
agreement of the data with elements that lack this signature
feature. If the deep \xm{} tail is missing in the data distributions
one must carefully assess that it is missing due to an astrophysical
feature in the spectrum, or it is missing due to acceptance of the
detector. As energy grows \xm{} increases and the possibility
increases that \xm{} can occur outside the field of view of the
detector, perhaps even in the ground for near vertical showers. Given
the relatively small exposure TA has for $E > 10^{19}$~eV, we cannot
reliably say either way yet. Given that starting at $10^{19}$~eV our
maximum likelihood tests result in simultaneous agreement with
multiple single species composition models, some with very large
differences in mass, we believe that we lack the statistical power to
draw firm conclusions about composition in this energy range. The
conclusions we draw from figures~\ref{fig:brlr_mxm} and
\ref{fig:brlr_hypothesis_test} is that BR/LR hybrid data is consistent
with a predominantly light composition below $10^{19}$~eV at the 95\%
confidence level, and more events must be collected above that energy
to draw further conclusions. These results do not imply that UHECR
composition is monospecific. Our tests were designed to be simple with
few free parameters.

\begin{figure}
  \centering
  \includegraphics[clip,width=0.98\linewidth]{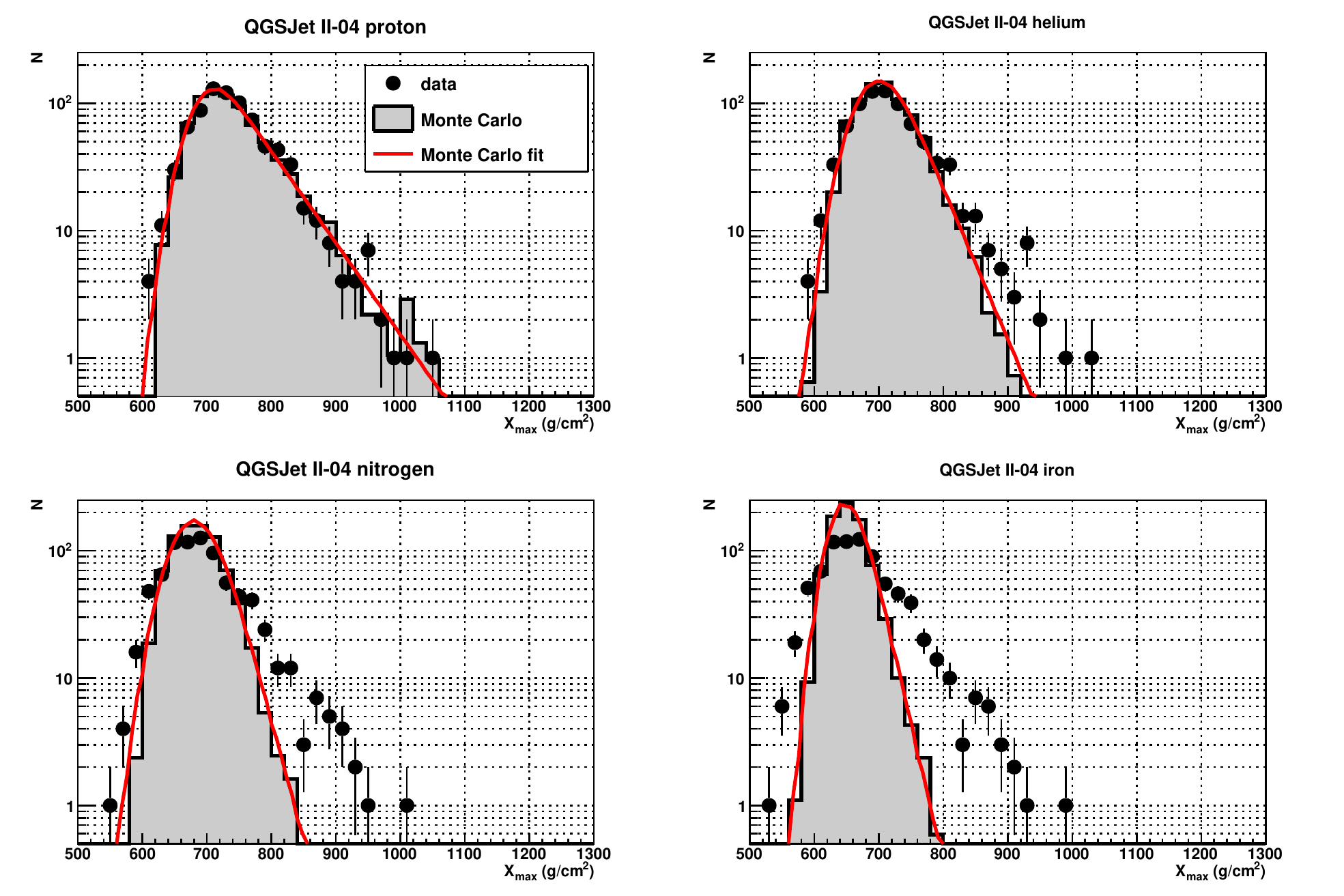}
  \caption{Comparison of BR/LR hybrid data and Monte Carlo \xm{}
    distributions after systematic shifting to maximize the likelihood
    function for $18.2 \leq \log_{10}(E/\mathrm{eV}) < 18.3$.}
    \label{fig:brlr_unbinnedML}
\end{figure}

Figure~\ref{fig:brlr_unbinnedML} shows the BR/LR hybrid data and Monte
Carlo \xm{} distributions after the maximum likelihood procedure has
been performed in one energy bin. Even though
figure~\ref{fig:brlr_mxm} indicates \mxm{} of data and QGSJET~II-04
protons differ, this figure does not account for the systematic
uncertainties in the data or in the model. After shifting,
figure~\ref{fig:brlr_unbinnedML} shows that the shapes of the data and
proton distributions agree quite well, especially in the tail of the
distributions. The same figure shows that even when shifting the same
data distribution to maximize the likelihood function for QGSJET~II-04
helium, the tails of the distribution don't agree as well as seen for
protons. This disagreement leads to a smaller probability (the
$p$-value) of observing the same likelihood recorded for the data
given the parent distribution is actually helium. The shapes of
QGSJET~II-04 nitrogen and iron have even worse agreement with the
shape of the data \xm{} distribution even after relatively large \xm{}
shifts are applied.

\begin{figure*}
  \centering
  \begin{subfigure}{0.46\linewidth}
    \includegraphics[clip,width=\textwidth]{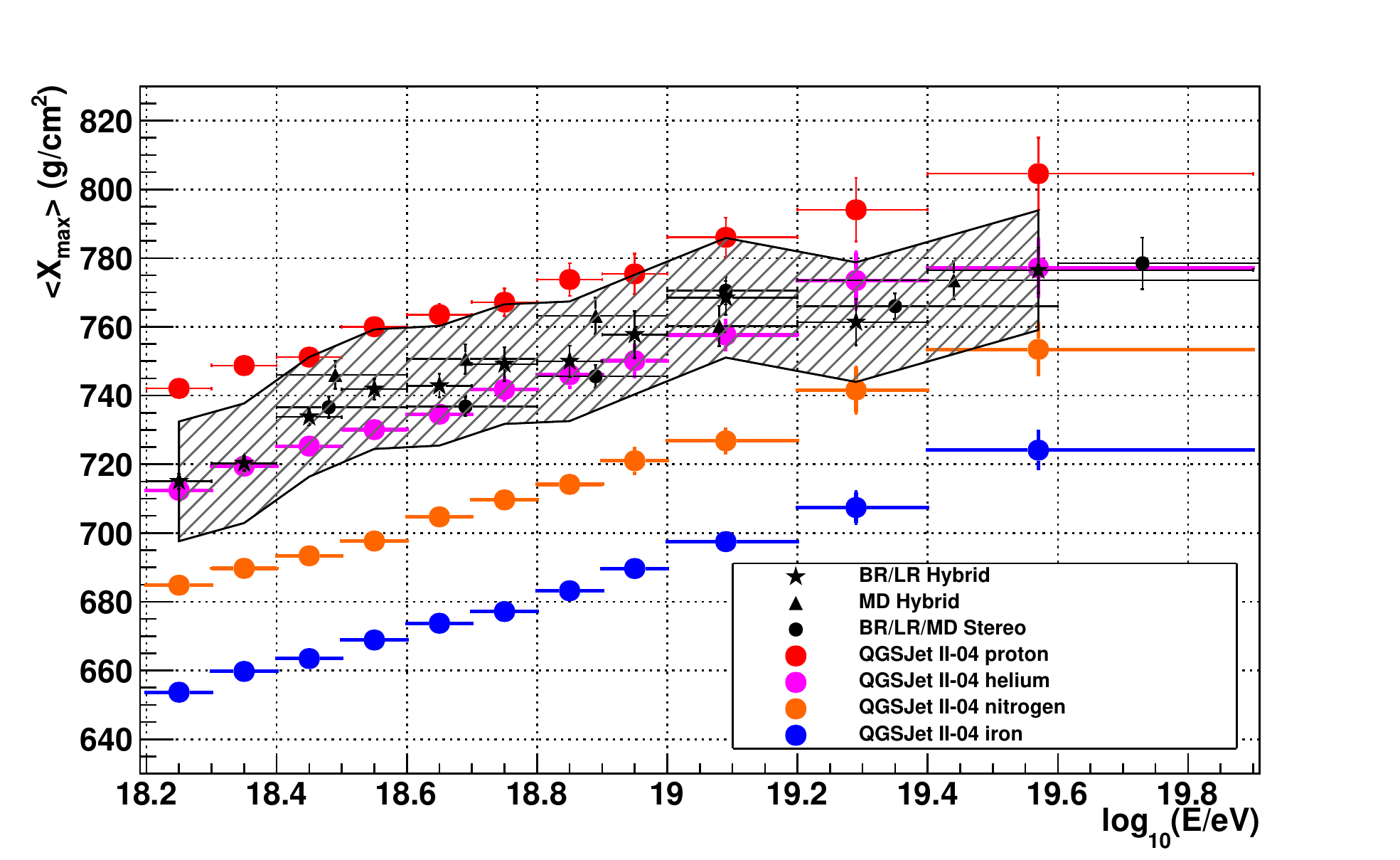}
    \caption{\mxm{} for three different TA composition
      measurements. }
    \label{fig:all_ta_mxm}
  \end{subfigure}%
  \qquad%
  \begin{subfigure}{0.46\linewidth}
    \includegraphics[clip,width=\textwidth]{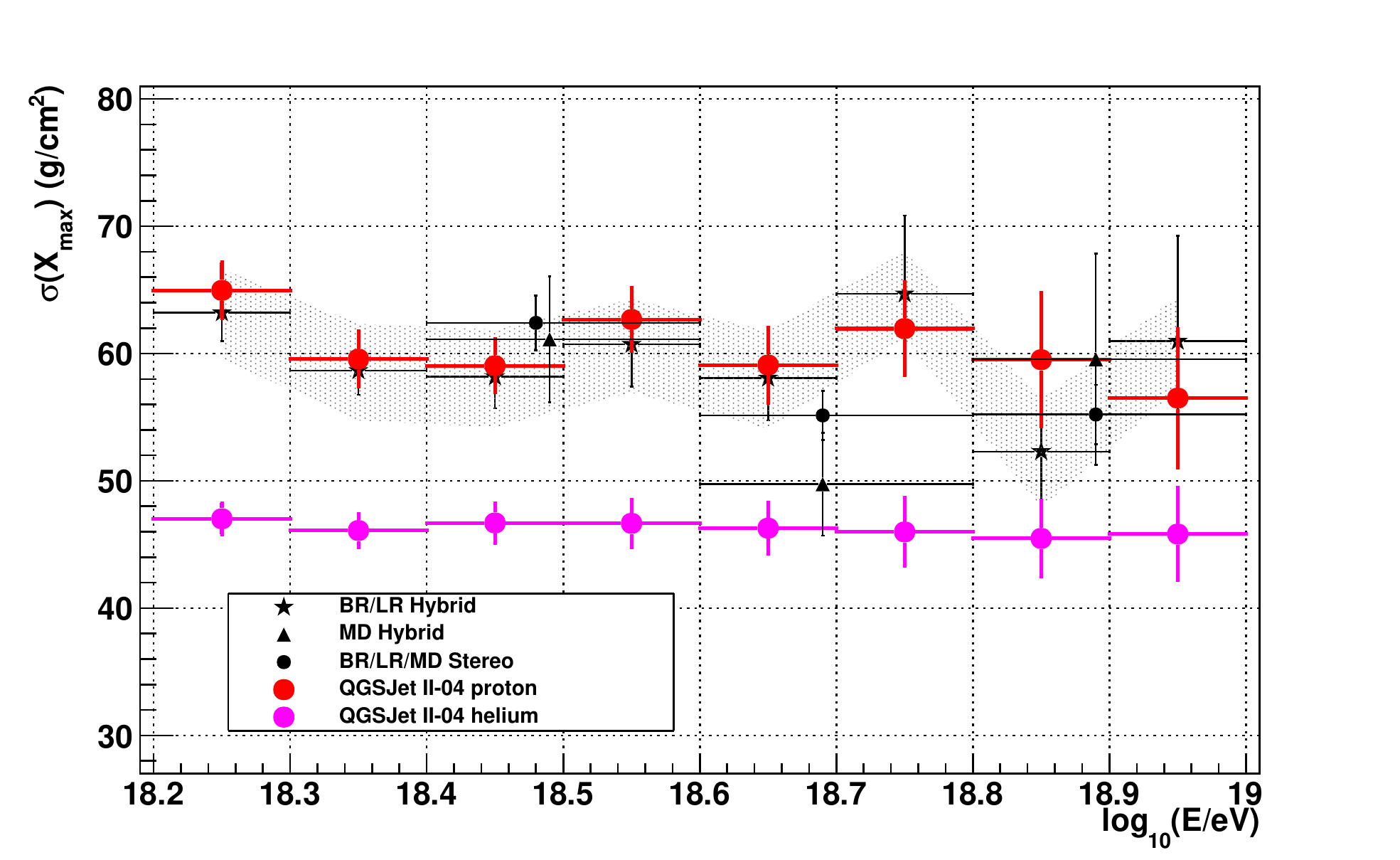}
    \caption{\sxm{} for three different TA composition measurements.}
    \label{fig:all_ta_sxm}
  \end{subfigure}
  \caption{\mxm{} and \sxm{} for all TA composition measurements that
    observe air shower \xm. QGSJET~II-04 predictions are for BR/LR
    hybrid reconstruction.}
  \label{fig:all_ta_moments}
\end{figure*}

If we compare the data of all TA FD-based analyses, they look
consistent within systematic
uncertainties. Figure~\ref{fig:all_ta_mxm} shows the observed data of
Middle Drum hybrid, TA FD stereo, and BR/LR hybrid, along with
predictions of BR/LR hybrid for four chemical
elements. Figure~\ref{fig:all_ta_sxm} shows \sxm{} for the same
measurements as well as the BR/LR hybrid predictions of QGSJET~II-04
proton and helium. We conclude that all three of these analyses are
consistent with a predominantly light composition.

\section{SD Composition Measurement}\label{sec:sd}
\begin{figure}
  \centering
  \includegraphics[clip,width=0.98\linewidth]{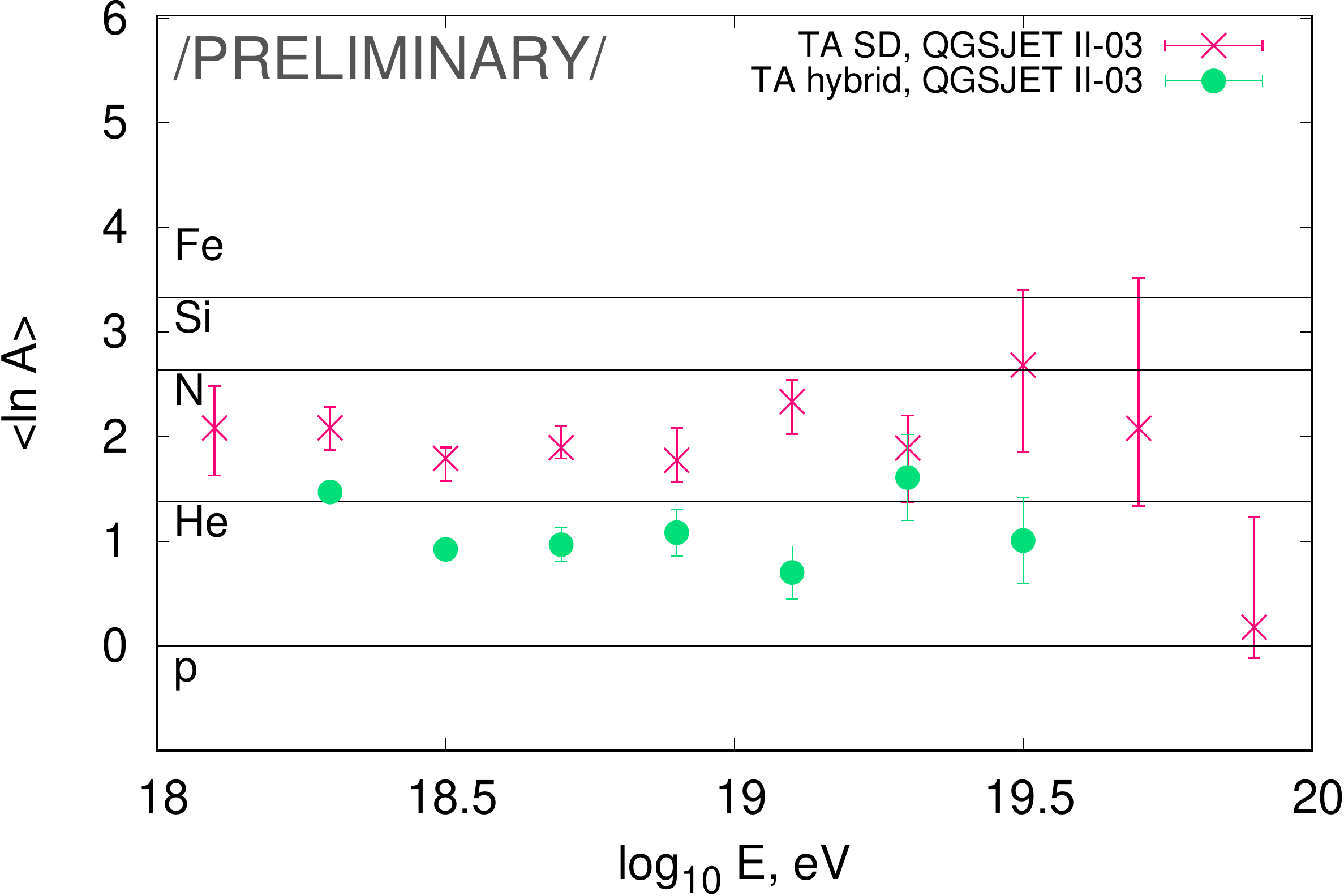}
  \caption{$\left<\ln A\right>$ observed by TA SD array using MVA
    analysis. $\left<\ln A\right>$ observed by BR/LR hybrid
    composition measurement is also shown. Both are in agreement with
    predominantly light composition.}
  \label{fig:sd_lnA}
\end{figure}

The strength of hybrid analysis is that it actually observes the depth
of \xm{}, along with recording the development of the shower before
and after, resulting in a very reliable measurement of \xm{} if the
shower geometry is sufficiently constrained. Methods that measure
\xm{} via air fluorescence have a low duty because of the requirement
to run during clear, moonless nights. SDs do not suffer from this
requirement, providing 100\% on time. If a method can be developed to
measure UHECR mass by SD array, then the gain in statistical power
will be enormous. TA has undertaken such an analysis which attempts to
classify the mass of the primary on an event-by-event basis using a
multivariate boosted decision tree technique. This analysis uses 14
composition sensitive variables collected by SD event analysis to
assign a continuous value, $\xi \in [-1:1]$, which classifies an event
as pure signal, $\xi = 1$, pure background, $\xi = -1$, or some value in
between the two. In this analysis, background is based upon pure
QGSJET~II-03 proton composition, and signal is pure QGSJET~II-03 iron
composition. The variables used in the multivariate analysis (MVA),
such as area-over-peak of the SD waveforms at 1200~m, shower front
curvature parameter, signal asymmetry in SD upper and lower layers,
are used to train a boosted decision tree (BDT) classifier using
proton and iron Monte Carlo. The classifier is then applied to the
data to calculate $\xi$ for each event. Using this information
$\left<\ln A\right>$ is calculated for the
data. Figure~\ref{fig:sd_lnA} shows the observed $\left<\ln A\right>$
using the MVA technique, as well as the predicted $\left<\ln A\right>$
using the BR/LR hybrid data. Using nine years of SD data, this
analysis collects 18007 events for $E \ge 10^{18}$~eV, about six times
the statistical power of the BR/LR hybrid analysis. The figure
indicates that the SD analysis finds TA data compatible with a
predominantly light composition for all energies above
$10^{18}$~eV. Under the assumption of a flat prior probability of an
equally weighted four component mixture of QGSJET~II-03 proton,
helium, nitrogen, and iron (fraction = 0.25 for each element), the
calculated systematic uncertainty of this analysis is $\delta
\left<\ln A\right> = 0.36$. See \cite{Abbasi:2018wlq} for further
details about this analysis.

\section{Summary}\label{sec:summary}
Telescope Array has been operating for ten years and has completed
composition analyses using four quasi-independent techniques: two
hybrid analyses using the SD array in conjunction with the Middle Drum
FD station and the combined Black Rock Mesa and Long Ridge FD
stations, stereo FD analysis using events observed only by two or more
FD stations, and a SD-only analysis using machine learning
algorithms. All four analyses observe UHECR composition consistent
with a predominantly light elements for $E \gtrsim 10^{18}$~eV. Our
highest statistics measurement of \xm, BR/LR hybrid, has limited
statistical power above $10^{19}$~eV, and we cannot draw reliable
conclusions about composition in this energy range. Comparison of
\mxm{} and \sxm{} of the three hybrid analyses show very consistent
results over all energies.

Because TA has sufficient exposure below $10^{19}$~eV to make
meaningful measurements of the shapes of the \xm{} distributions we
observe, and model dependence and systematic uncertainty of hadronic
models is an issue for UHECR composition, we have developed methods to
test our data distributions against simulations, not just relying on
the first and second moments. These methods leverage the additional
information that comes from using the entire set of data in a given
energy bin, by calculating the probability of observing our data given
that it was generated according to some model. These are
non-parametric, distribution free tests which test the shapes of
distributions in question and allow for systematic uncertainty in the
data or models. These tests allow us to empirically measure the
agreement of data with simulations at whatever confidence level we
choose.

The SD-only analysis using MVA is a promising measurement which does
not use the traditional method of observing \xm{} to infer
composition, but has vastly greater statistical power. It also
measures light composition and compares well with the predicted
$\left<\ln A\right>$ of the BR/LR hybrid analysis.

TA will continue collect data with the planned TAx4 upgrade already
underway. This will improve TA's exposure above $10^{19}$~eV,
providing needed statistics to allow us to firm up our measurements in
this energy range.

\section*{Acknowledgements}
The Telescope Array experiment is supported by the Japan Society for
the Promotion of Science (JSPS) through 
Grants-in-Aid
for Priority Area
%"Highest Energy Cosmic Rays"
431,
for Specially Promoted Research 
%``Extreme Phenomena in the Universe Explored by Highest Energy Cosmic Rays'' 
%Grant Number 
JP21000002, 
%Grant-in-Aid 
for Scientific  Research (S) 
%"Quest for the unified picture of the explosion mechanism of supernovae and the central engine of gamma-ray bursts"
%Grant Number 
JP19104006, 
%Grant-in-Aid 
for Specially Promoted Research 
%"Extended Telescope Array Experiment - Nearby Extreme Universe Elucidated by Highest-energy Cosmic Rays"
%Grant Number 
JP15H05693, 
%Grant-in-Aid 
for Scientific  Research (S)
%"Study of the ultra high energy cosmic ray source evolution by detailed measurement of cosmic rays in the wide energy range"
%Grant Number 
JP15H05741 and
%Grant-in-Aid 
for Young Scientists (A)
%"hoge hoge"
%Grant Number 
JPH26707011; 
by the joint research program of the Institute for Cosmic Ray Research (ICRR), The University of Tokyo; 
by the U.S. National Science
Foundation awards PHY-0601915,
PHY-1404495, PHY-1404502, and PHY-1607727; 
by the National Research Foundation of Korea
% \linebreak
(2016R1A2B4014967, 2016R1A5A1013277, 2017K1A4A3015188, 2017R1A2A1A05071429) ;
%\linebreak 
by the Russian Academy of
Sciences, RFBR grant 16-02-00962a (INR), IISN project No. 4.4502.13,
and Belgian Science Policy under IUAP VII/37 (ULB). The foundations of
Dr. Ezekiel R. and Edna Wattis Dumke, Willard L. Eccles, and George
S. and Dolores Dor\'e Eccles all helped with generous donations. The
State of Utah supported the project through its Economic Development
Board, and the University of Utah through the Office of the Vice
President for Research. The experimental site became available through
the cooperation of the Utah School and Institutional Trust Lands
Administration (SITLA), U.S. Bureau of Land Management (BLM), and the
U.S. Air Force. We appreciate the assistance of the State of Utah and
Fillmore offices of the BLM in crafting the Plan of Development for
the site.  Patrick Shea assisted the collaboration with valuable advice 
on a variety of topics. The people and the officials of Millard County, 
Utah have been a source of
steadfast and warm support for our work which we greatly appreciate. 
We are indebted to the Millard County Road Department for their efforts 
to maintain and clear the roads which get us to our sites. 
We gratefully acknowledge the contribution from the technical staffs of
our home institutions. An allocation of computer time from the Center
for High Performance Computing at the University of Utah is gratefully
acknowledged.

\bibliography{HANLON_William_UHECR2018}

\end{document}